\documentclass[11pt]{article}
\usepackage{moriond,epsfig}

\usepackage{longtable}

\newcommand{\mb}{\ensuremath{M_{\mathrm{bc}}}}

\newcommand{\etapi}{\ensuremath{\eta_{\pi^+ \pi^- \pi^0}}}
\newcommand{\btoetak}{\ensuremath{B^{\pm} \to \eta K^{\pm}}}
\newcommand{\btoetapi}{\ensuremath{B^{\pm} \to \eta \pi^{\pm}}}

\newcommand*{\brkspp}{53.2\pm11.3\pm9.7}
\newcommand*{\brkskk}{34.8\pm6.7\pm6.5}
\newcommand*{\brkskp}{9.3}

\newcommand*{\brkcpp}{55.6\pm5.8\pm7.7}
\newcommand*{\brkckk}{35.3\pm3.7\pm4.3}
\newcommand*{\brkckp}{12}

\newcommand*{\brksppkstpi}{13.5^{+5.0}_{-4.4}\pm{2.9}}
\newcommand*{\brksppkstxpi}{22.9^{+8.7}_{-8.0}\pm{6.0}}
\newcommand*{\brkspprhoks}{12.4}
\newcommand*{\brksppfzeroks}{14.2}
\newcommand*{\brksppfxks}{13.7}

\newcommand*{\brkcppkstpi}{12.9^{+2.8+1.4+2.3}_{-2.6-1.4-4.5}}
\newcommand*{\brkcppkstxpi}{14.5^{+3.5+1.8+3.3}_{-3.3-1.8-6.5}}
\newcommand*{\brkcpprhokc}{12}
\newcommand*{\brkcppfzerokc}{9.6^{+2.5+1.5+3.4}_{-2.3-1.5-0.8}}
\newcommand*{\brkcppfxkc}{11.1^{+3.4+1.4+7.2}_{-3.1-1.4-2.9}}

\newcommand*{\brkskkphiks}{6.4^{+3.0}_{-2.6}\pm1.3}
\newcommand*{\brkskkfxks}{20.4^{+5.3}_{-4.9}\pm3.8}
\newcommand*{\brkskkak}{12.1}
\newcommand*{\brkskkaxk}{10.0}

\newcommand*{\brkckkphikc}{7.2^{+1.5+0.9+0.4}_{-1.4-0.9-0.4}}
\newcommand*{\brkckkfxkc}{27.6^{+3.2+3.5+1.4}_{-3.2-3.5-1.4}}

\newcommand{\pppi}{\ensuremath{ p \overline{p} \pi^{\pm} }}
\newcommand{\ppks}{\ensuremath{ p \overline{p} K^0_S }}

\def\be{\begin{equation}}
\def\ee{\end{equation}}
\def\bea{\begin{eqnarray}}
\def\eea{\end{eqnarray}}

\begin{document}
\vspace*{4cm}
\title{RARE \boldmath $B$ DECAYS AT BELLE}

\author{ H.C. HUANG \\ (for the Belle Collaboration) }

\address{Department of Physics, National Taiwan University, Taipei
106, Taiwan}

\maketitle\abstracts{
Recent results of rare $B$ decay analyses based on 31.9 
million $B\bar{B}$ collected with the Belle detector at the KEKB asymmetric 
$e^+e^-$ collider  are presented.  
We have made the first observation of charmless baryonic decay $B^\pm
\to p \overline{p} K^\pm$, the three-body $B^0 \to K^0 \pi^+ \pi^-$
and $B^0 \to K^0 K^+ K^-$.  The measured branching fractions are 
$\mathcal{B}(B^+\to p \bar{p} K^+) =
(4.3^{+1.1}_{-0.9} \pm 0.5)\times 10^{-6}$ 
, $\mathcal{B}(B^0{\to}K^0\pi^+\pi^-) = (\brkspp) \times
10^{-6}$, and ${\mathcal{B}}(B^0{\to}K^0 K^+ K^-) = (\brkskk)
\times 10^{-6}$. 
We also see strong evidence of $B^\pm \to
\eta K^\pm$ and $B^\pm \to \eta \pi^\pm$, and observed the decay
$B^\pm \to \omega K^\pm$ with 
$\mathcal{B}(B^\pm \to \omega K^\pm) = (9.9^{+2.7}_{-2.4} \pm 1.0)
\times 10^{-6}$.  
Preliminary results of improved
measurements of the branching 
fractions for the decays $B \to K \pi$ and $\pi \pi$ are reported.  
No evidence for direct $CP$ violation is found in the decays $B^\pm \to K^\pm
\pi^\mp$, $K^\pm \pi^0$, $K^0 \pi^\pm$, $\pi^\pm \pi^0$, and $\omega
K^\pm$.
}

%%%%%%%%%%%%%%%%%%%%%%%%%%%%%%%%%%%%%%%%%%%%%%%%%%%%%%%%%%%
\section{Introduction}
%%%%%%%%%%%%%%%%%%%%%%%%%%%%%%%%%%%%%%%%%%%%%%%%%%%%%%%%%%%

One of the most important goals of experiments at B-factories is to precisely 
measure the sides and angles of the unitarity triangle in the 
Cabbibo-Kobayashi-Maskawa matrix~\cite{ckm} (CKM) and check its
consistency.  Any inconsistency is a clear signal of new physics  
beyond the Standard Model (SM).  Rare $B$ decays play an important
role in working towards this goal. 

Preliminary results of rare $B$ decay analyses based on a 
31.9 million $B\bar{B}$ sample are presented here.  The data were 
collected with the Belle detector~\cite{REF:belle} at the KEKB asymmetric 
$e^+e^-$ collider.~\cite{REF:kekb}

%%%%%%%%%%%%%%%%%%%%%%%%%%%%%%%%%%%%%%%%%%%%%%%%%%%%%%%%%%%%%%%%%%%%%%
%\section{Belle Detector}
%%%%%%%%%%%%%%%%%%%%%%%%%%%%%%%%%%%%%%%%%%%%%%%%%%%%%%%%%%%%%%%%%%%%%%

Belle is a general-purpose detector with a 1.5 T superconducting
solenoid magnet.  
Charged particle tracking, covering 92\% of the total center-of-mass (CM) 
solid angle, is provided by the Silicon Vertex Detector (SVD) consisting of
three concentric layers of double-sided silicon strip
detectors and a 50-layer Central Drift Chamber (CDC). 
Charged hadrons are distinguished by combining the responses
from an array of Silica Aerogel \v Cerenkov Counters (ACC), 
a Time of Flight Counter system (TOF), and $dE/dx$ measurements in the CDC. 
The combined response provides $K/\pi$ separation of at least 2.5$\sigma$
for laboratory momenta up to 3.5 GeV/$c$.  
Photons and electrons are detected in an array of 8736 
%CsI(T$\ell$) 
CsI(Tl) 
crystals (ECL) located inside the magnetic field and
covering the entire solid angle of the charged particle tracking system.  
The 1.5~T magnetic field is returned via an iron
yoke, instrumented to detect muons and $K_L$ mesons (KLM).
The KLM consists of alternating layers of resistive plate chambers and
4.7 cm thick steel plates.

%%%%%%%%%%%%%%%%%%%%%%%%%%%%%%%%%%%%%%%%%%%%%%%%%%%%%%%%%%%
%\section{Analysis in General}
%%%%%%%%%%%%%%%%%%%%%%%%%%%%%%%%%%%%%%%%%%%%%%%%%%%%%%%%%%%

In all the decay modes presented here, the continuum process 
($e^+e^- \to q\bar{q}$) is the dominant background. 
Since $B\bar{B}$ events are spherical while the continuum events are 
jet-like, 
we apply cuts on various event shape variables (such as sphericity, thrust 
angle, Fox-Wolfram moments, and the production angle of $B$) to suppress the 
background.

 $B$ candidates are identified using two kinematic variables: beam 
constrained mass: $\mb =\sqrt{E^{2}_{beam} - p^{2}_B}$, and the energy 
difference: $\Delta E = E_B - E_{beam}$. Here $E_{beam}$ is the beam 
energy, $p_B$ and $E_B$ are the momentum and energy of a reconstructed 
$B$ candidate, respectively, where all variables are defined in 
the $\Upsilon(4S)$ rest frame. 

$K/\pi$ separation is performed by applying a cut on the likelihood ratio,
$L_K/(L_\pi + L_K)$, where $L_K~(L_\pi)$ is a kaon (pion) likelihood 
computed from information from the particle identification devices:
specific ionization loss in the central drift chamber, photo-electron yield 
in the aerogel Cherenkov counters, and time-of-flight.~\cite{REF:belle}

%%%%%%%%%%%%%%%%%%%%%%%%%%%%%%%%%%%%%%%%%%%%%%%%%%%%%%%%%%%%%%%%%%%%%%
\section{Two-Body Charmless \boldmath Hadronic $B$ Decays}
%%%%%%%%%%%%%%%%%%%%%%%%%%%%%%%%%%%%%%%%%%%%%%%%%%%%%%%%%%%%%%%%%%%%%%

These processes are manifestations of penguin or suppressed three
amplitudes proportional to small couplings in hadronic flavor mixing
(CKM matrix).  Because of the
absence of CKM favored $b \to c$ amplitudes, these decays are
particularly sensitive to potentially new contributions from
interference effects and virtual particles in loops.

\begin{table}
\caption{The signal yield, reconstruction efficiency ($\epsilon$), 
statistical significance ($\Sigma$), branching fractions ($\mathcal{B}$), 
and the $90\%$ confidence level upper limits (UL) for two-body $B$
decay modes.  All the results are preliminary.
\label{tb:2body}}
\vspace{0.4cm}
\small
\begin{center}
\begin{tabular}{|l|ccc|cc|}
\hline		
Mode	& Yield	& $\epsilon$
	& $\Sigma$ & $\mathcal{B}$ ($\times 10^{-6}$) & UL ($\times
	10^{-6}$)
\\
\hline
$K^+\pi^-$&
$218 \pm 18$&
0.31&
16.4&
{ $21.8 \pm 1.8 \pm 1.5$}
&-
\\
$K^+\pi^0$&
$58 \pm 11$&
0.15&
6.3&
{ $12.5 \pm 2.4 \pm 1.2$}
&-
\\
$K^0\pi^+$&
$66 \pm 10$&
0.32&
8.2&
{ $18.8 \pm 3.0 \pm 1.5$}
&-
\\
$K^0\pi^0$&
$19 \pm 8$&
0.23&
2.7&
{ $7.7 \pm 3.2 \pm 1.6$}
&-
\\
\hline
$\pi^+\pi^-$&
$51 \pm 11$&
0.31&
5.4&
{ $5.1 \pm 1.1 \pm 0.4$}
&-
\\
$\pi^+\pi^0$&
$36 \pm 11$&
0.16&
3.5&
{ $7.0 \pm 2.2 \pm 0.8$}&
-
\\
$\pi^0\pi^0$&
$12 \pm 6$&
0.13&
2.2&
-&
{ $< 5.6$}
\\
\hline
$K^+K^-$&
$0\pm 2$&
0.26&
0&
-&
{ $< 0.5$}
\\
$K^+K^0$&
$0 \pm 2$&
0.17&
0&
-&
{ $< 3.8$}
\\
$K^0K^0$&
$1 \pm  3$&
0.20&
0&
-&
{ $< 13$}
\\ \hline
$\eta K^+$
	& -
	& -
	& {4.9}
	& {$5.3 ^{+1.8}_{-1.5} \pm 0.6$}
	& - 
\\
~~$\eta_{\gamma \gamma} K^{+} $
	& $12.7 ^{+5.0}_{-4.2}$
	& 0.18
	& 4.3 
	& $5.7 ^{+2.2}_{-1.9}$
	& - 
\\
~~$\eta_{\pi^{+} \pi^{-} \pi^{0}} K^{+} $
	& $ 4.2 ^{+3.1}_{-2.3} $
	& 0.15
	& 2.4 
	& $4.6 ^{+3.4}_{-2.5}$
	& - 
\\ \hline
$\eta \pi^+$
	& -
	& -
	& {4.3}
	& {$5.4 ^{+2.0}_{-1.7} \pm 0.6$}
	& - 
\\
~~$\eta_{\gamma \gamma} \pi^{+} $  
	& $11.4 ^{+4.9}_{-4.1}$
	& 0.16
	& 3.8
	& $5.9 ^{+2.5}_{-2.1}$
	& - 
\\
~~$\eta_{\pi^{+} \pi^{-} \pi^{0}} \pi^{+} $
	& $ 4.0 ^{+3.1}_{-2.3}$	
	& 0.14
	& 2.1
	& $4.8 ^{+3.7}_{-2.8}$
	& - 
\\
\hline
$\omega K^-$	& $19.7^{+5.4}_{-4.8}$  %{}^{+0.7}_{-0.5}$ 
		& $0.06$
		& 6.4
		& $9.9^{+2.7}_{-2.4} \pm 1.0$ 
		& - 
\\
$\omega \pi^-$	& $10.6^{+4.8}_{-4.5}$ %{}^{+0.4}_{-0.6}$ 
		& $0.08$
		& 3.3
		& $4.3^{+2.0}_{-1.8} \pm 0.5$
		& $< 8.2$
\\
\hline
\end{tabular}
\end{center}
\end{table}

%%%%%%%%%%%%%%%%%%%%%%%%%%%%%%%%%%%%%%%%%%%%%%%%%%%%%%%%%%%
\subsection{$B \to \pi\pi$, $K\pi$, $KK$}
%%%%%%%%%%%%%%%%%%%%%%%%%%%%%%%%%%%%%%%%%%%%%%%%%%%%%%%%%%%

The charmless hadronic $B$ decays $B \to \pi\pi$, $K \pi$, and $KK$
provide a rich sample to test the standard model and to probe new
physics.  Of particular interest are indirect and direct $CP$
violation in the $\pi \pi$ and $K \pi$ modes, which are related to the
angles $\phi_2$ and $\phi_3$ of the unitarity triangle,
respectively.~\cite{kpitheory}  
Measurements of branching fractions of these decay modes are
an important step towards these $CP$ violation studies. 

\begin{figure}
\begin{center}
\includegraphics[width=0.85in, trim=60 0 0 0]{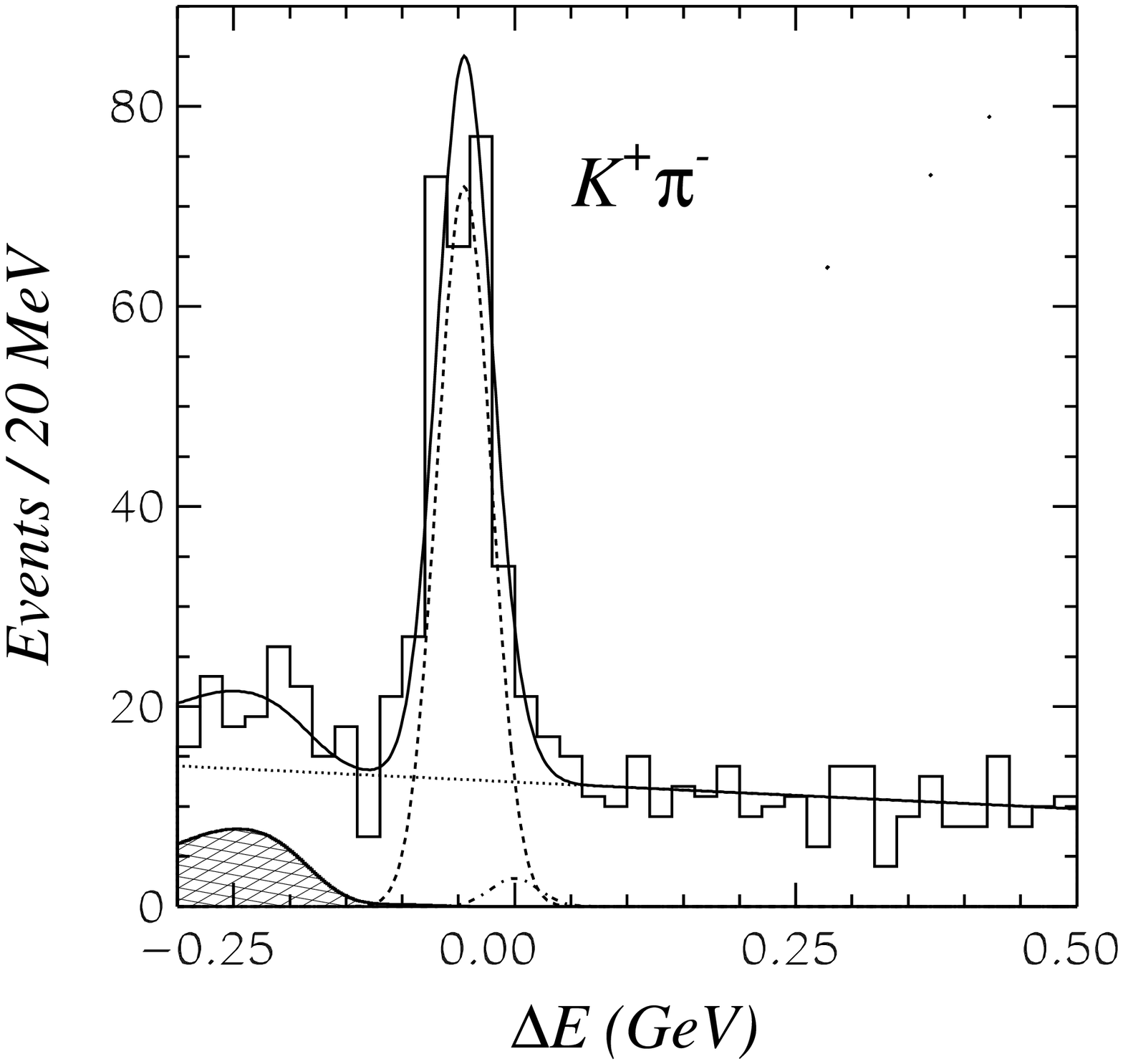}
\includegraphics[width=0.85in, trim=60 0 0 0]{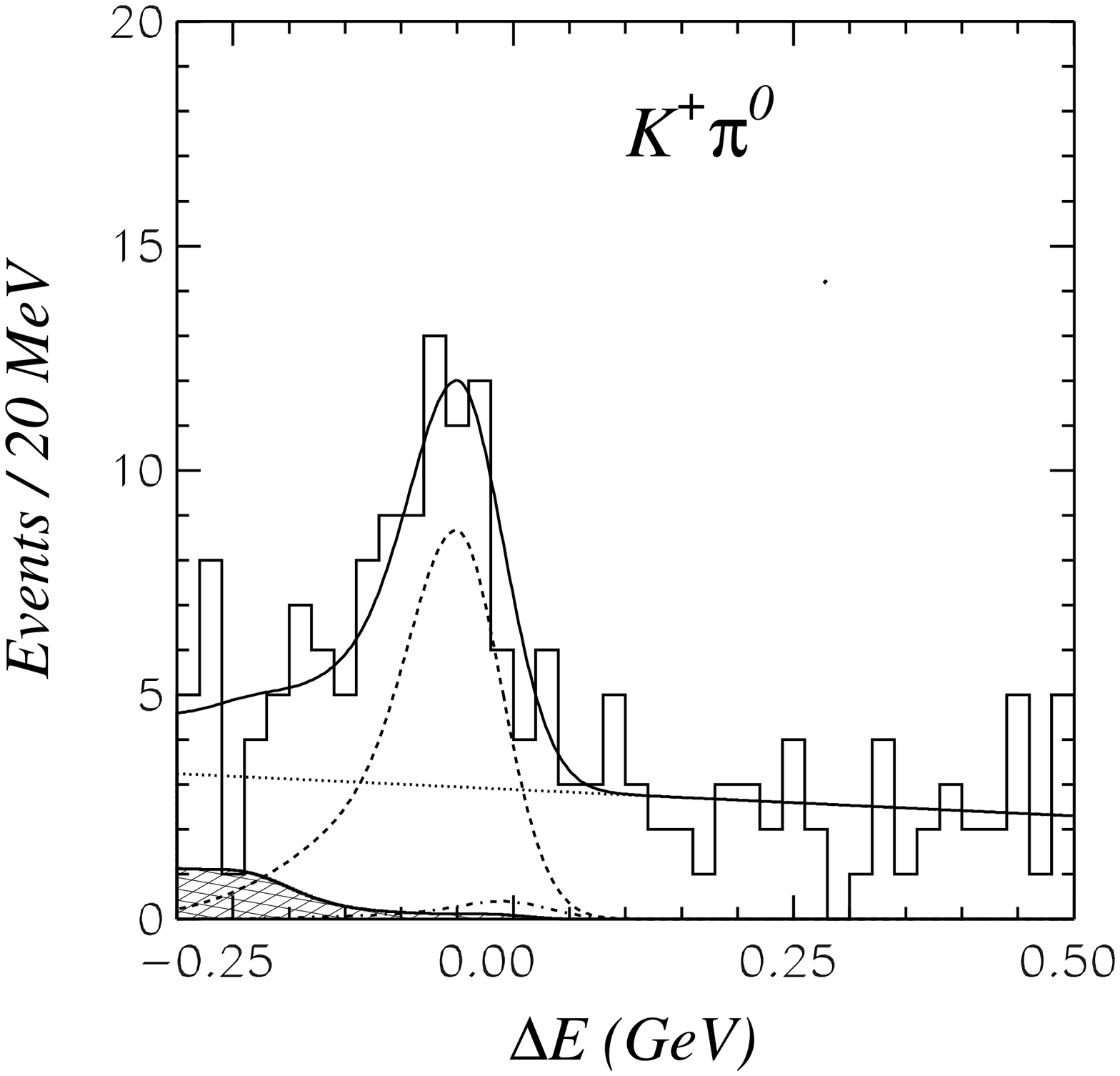}
\includegraphics[width=0.85in, trim=60 0 0 0]{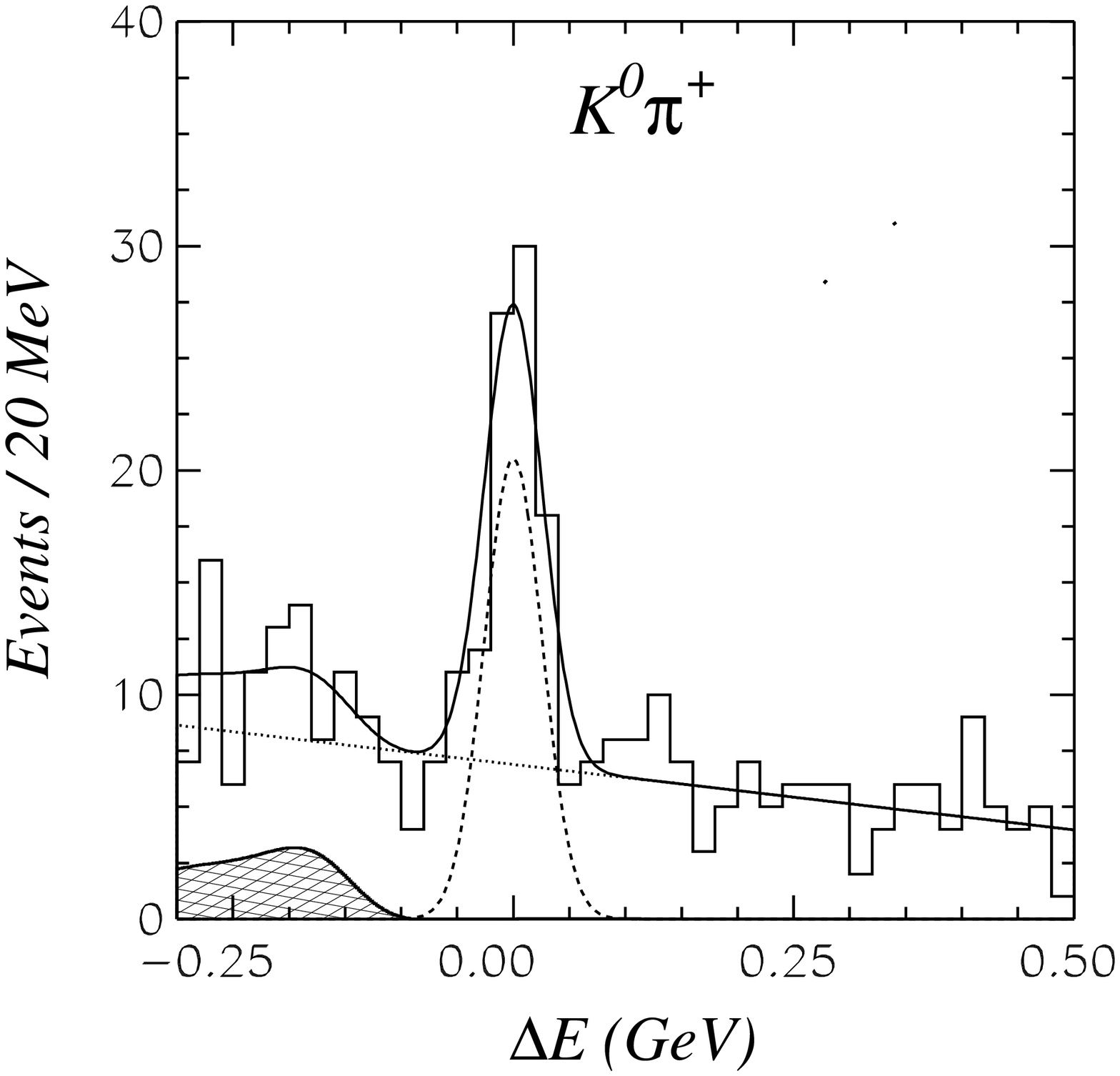}
\includegraphics[width=0.85in, trim=60 0 0 0]{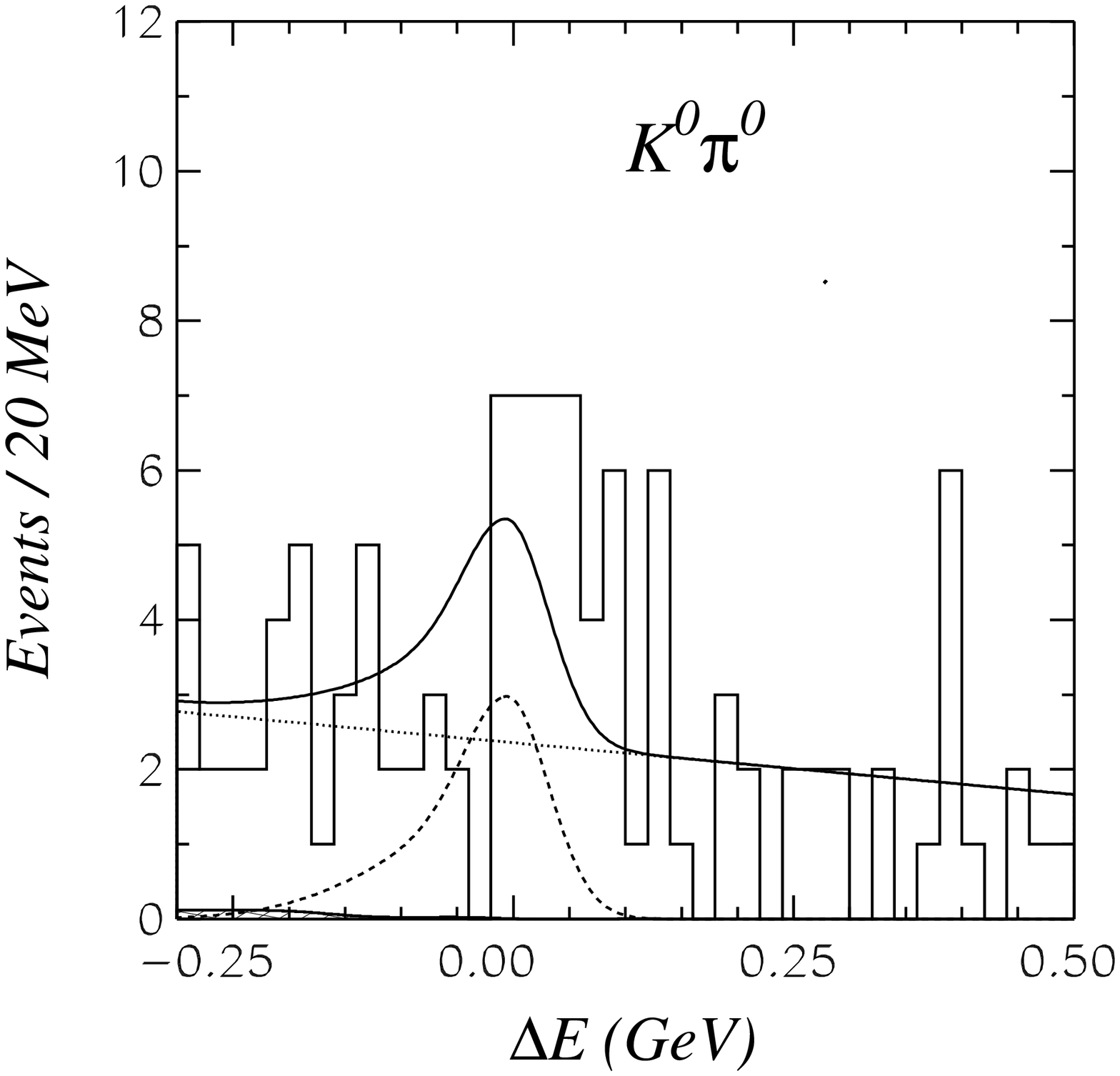}
\includegraphics[width=0.85in, trim=60 0 0 0]{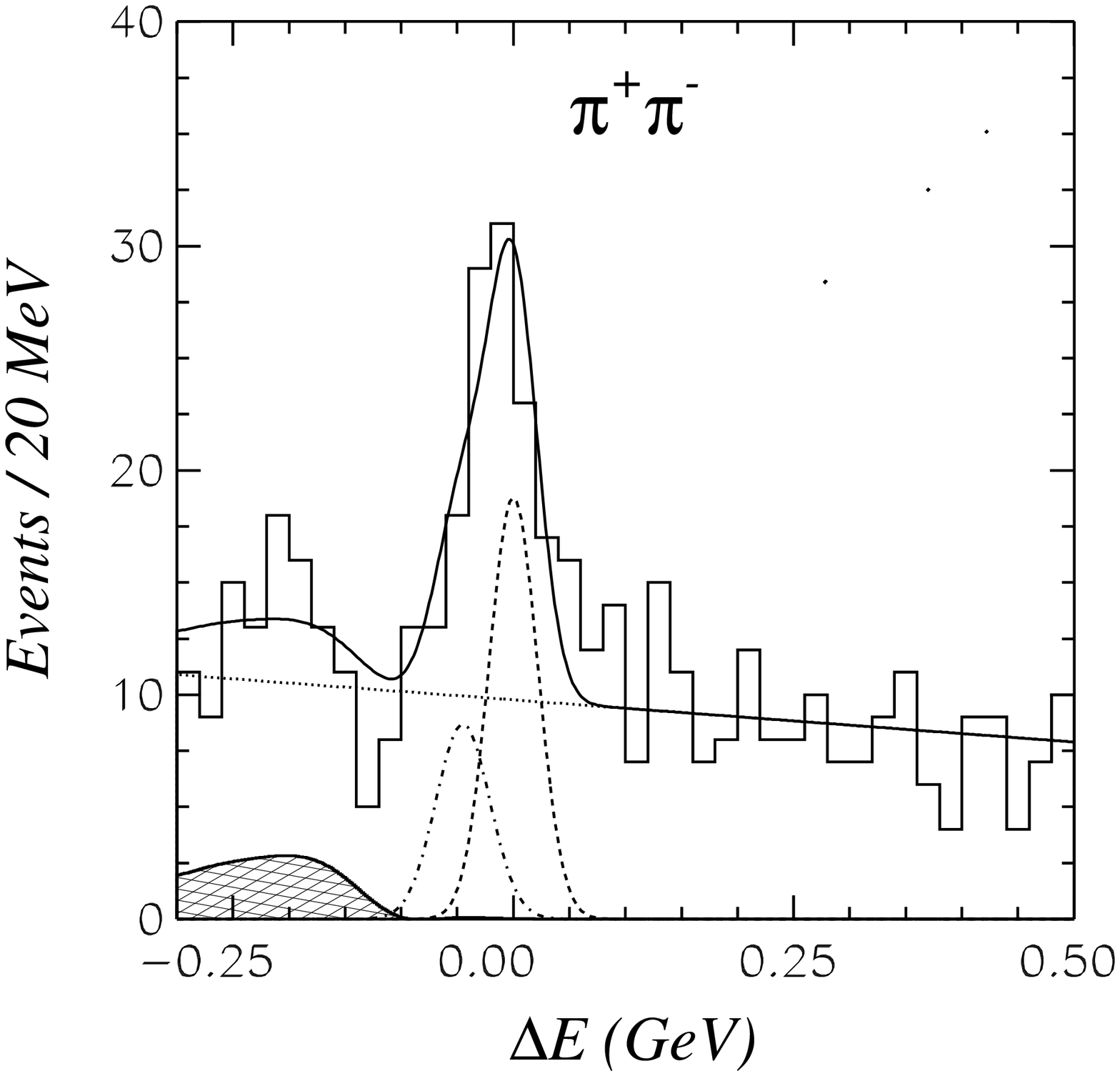}
\includegraphics[width=0.85in, trim=60 0 0 0]{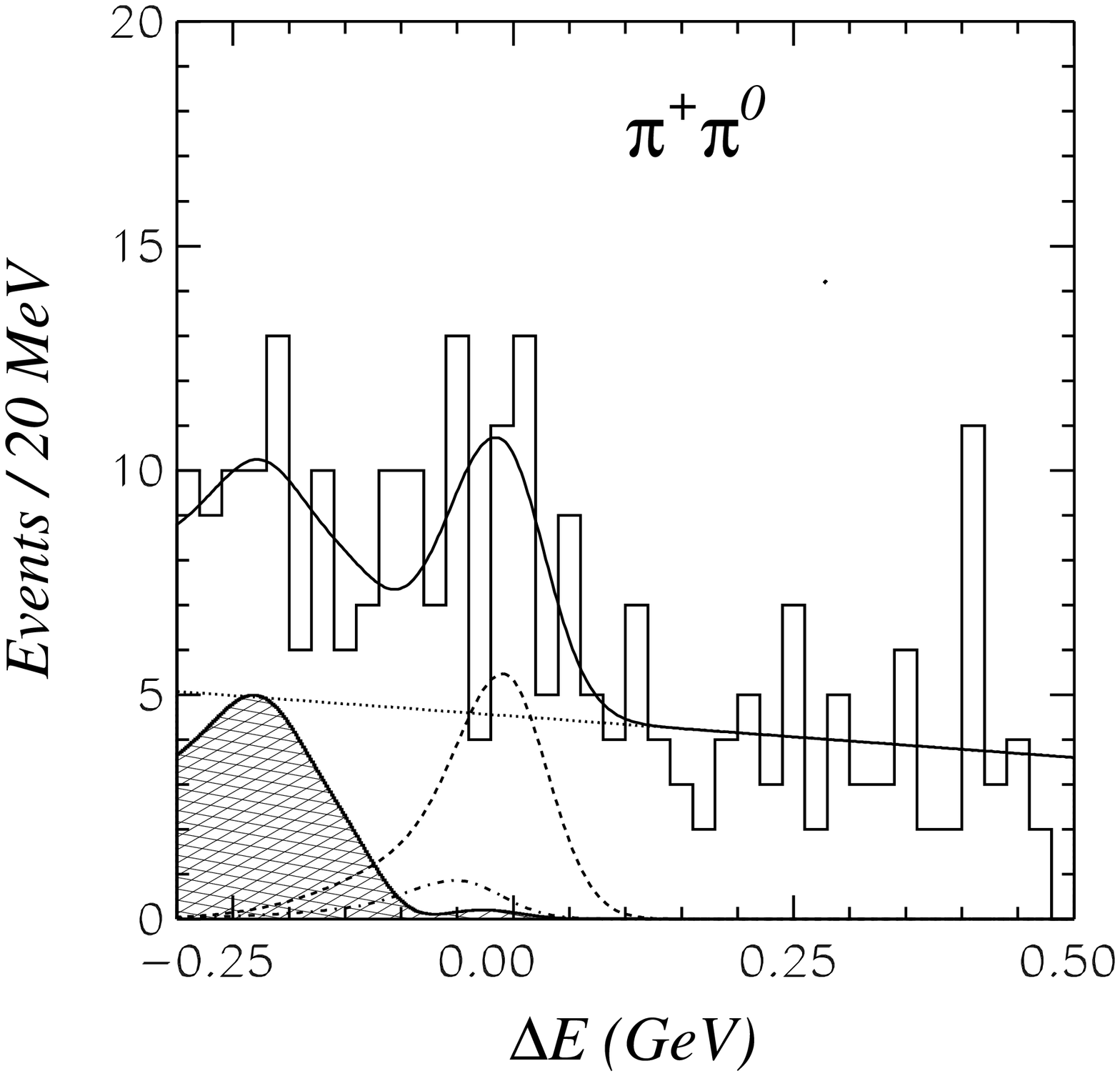}
\includegraphics[width=0.85in, trim=60 0 0 0]{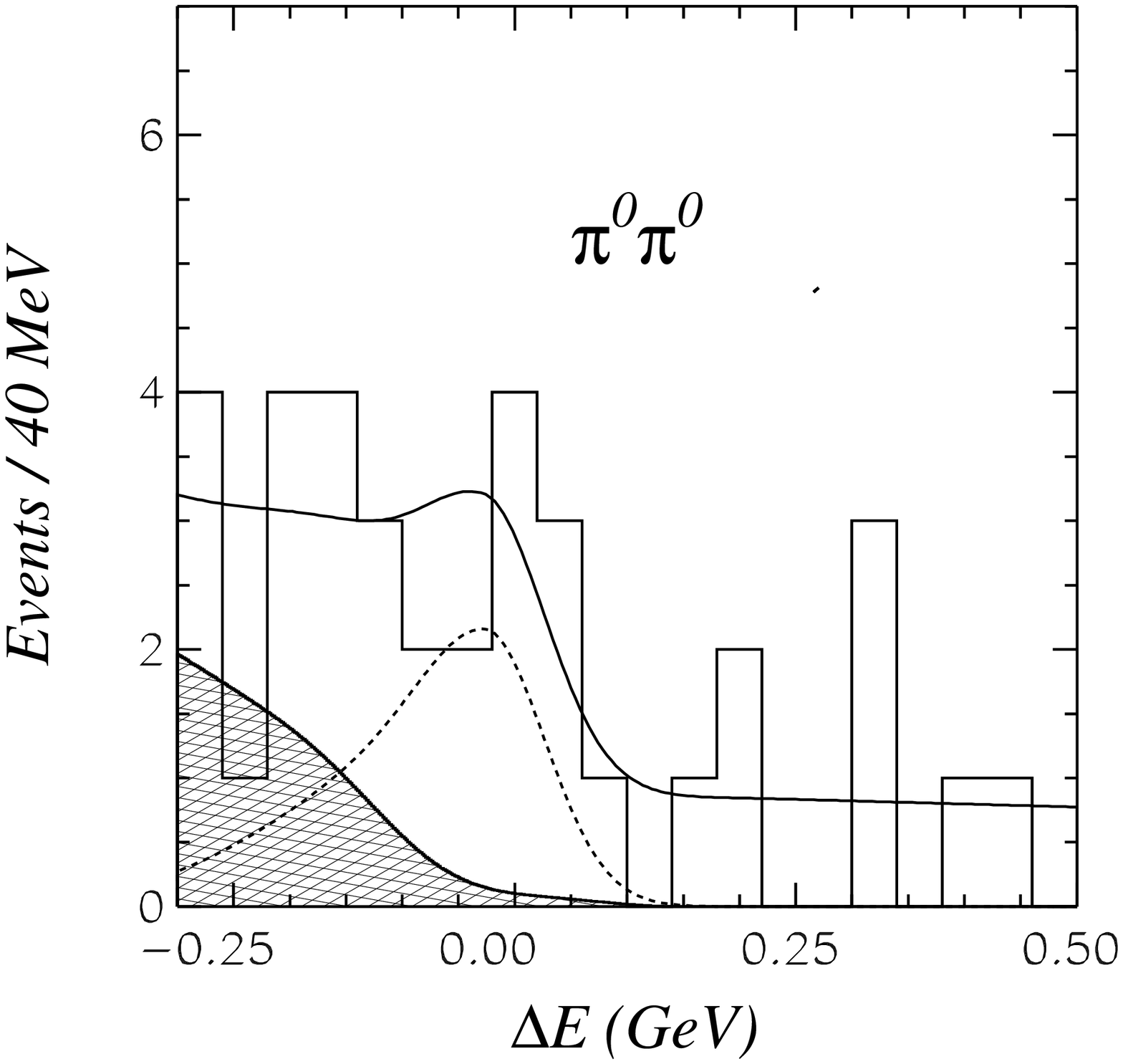}
\end{center}
\caption{$\Delta E$ distributions of $B \to K\pi$ and $\pi\pi$ decays:
(from left to right) $K^+ \pi^-$, $K^+ \pi^0$, $K^0_S \pi^+$, $K^0_S
\pi^0$, $\pi^+ \pi^-$, $\pi^+ \pi^0$, and $\pi^0 \pi^0$.
\label{fg:hh}}
\end{figure}

In this analysis, $B$ meson candidates are reconstructed in ten 
combinations: $h^+ h'^-$, $h^+ \pi^0$, $K^0_S h^+$, $K^0_S \pi^0$,
$K^0_S K^0_S$, and $\pi^0 \pi^0$, where the symbols $h$ and $h'$ refer
to $\pi$ or $K$.  Candidate 
$\pi^0$ mesons are formed from pairs of photons and 
candidate $K^0_S$ mesons are reconstructed from pairs of oppositely
charged tracks with a displaced vertex from the interaction point.  
Signal yields are extracted by fits to the $\Delta E$ distributions 
taking into account feed-across from other mis-identified $B \to hh'$
decays and backgrounds from multi-body and radiative charmless $B$
decays.  The fit results are given in Table~\ref{tb:2body} and the $\Delta
E$ distributions are shown in Figure~\ref{fg:hh}.

%%%%%%%%%%%%%%%%%%%%%%%%%%%%%%%%%%%%%%%%%%%%%%%%%%%%%%%%%%%%%%%%%%%%%%
\subsection{$B^\pm \to \eta h^\pm$}
%%%%%%%%%%%%%%%%%%%%%%%%%%%%%%%%%%%%%%%%%%%%%%%%%%%%%%%%%%%%%%%%%%%%%%

Previous measurements~\cite{cleoetakst,babaretapk} yielded large rates
for $B \to \eta' K$ 
and $B \to \eta K^*$, motivating a number of new theoretical ideas.  
Measurement of related decays $B \to \eta K$ can help clarify these.
Besides, it has been suggested that the decays of $B^+ \to \eta \pi^+$
is a good candidate for observing direct $CP$ violation.~\cite{rosner01} 

In this analysis, we reconstruct $\eta$ mesons using the $\eta \to
\gamma\gamma$ and $\eta\to \pi^+\pi^-\pi^0$ decay channels.
Candidate $\eta$ mesons are required to 
have invariant masses within $\pm 2.5\sigma$ of the $\eta$ peak, 
where $\sigma$ is $10.6~{\rm
MeV}/c^2$ and $3.4$ MeV/$c^2$ for the $\gamma\gamma$ and
$\pi^+\pi^-\pi^0$ modes, respectively. 
For the $\pi^+\pi^-\pi^0$ mode, 
the $\pi^+\pi^-$ pair is constrained to a vertex.
Both photons from the $\eta \to \gamma\gamma$ mode are 
required to have $E_{\gamma} > 100$ MeV and
we remove $\eta$ candidates if either of the daughter photons
can be combined with any other photon with 
$E_{\gamma} > 100$ MeV to form a $\pi^0$ candidate.
The $\eta$ candidates are further constrained to the known $\eta$
mass.~\cite{pdg}

\begin{figure} [b]
\begin{center}
\includegraphics[width=1.4in]{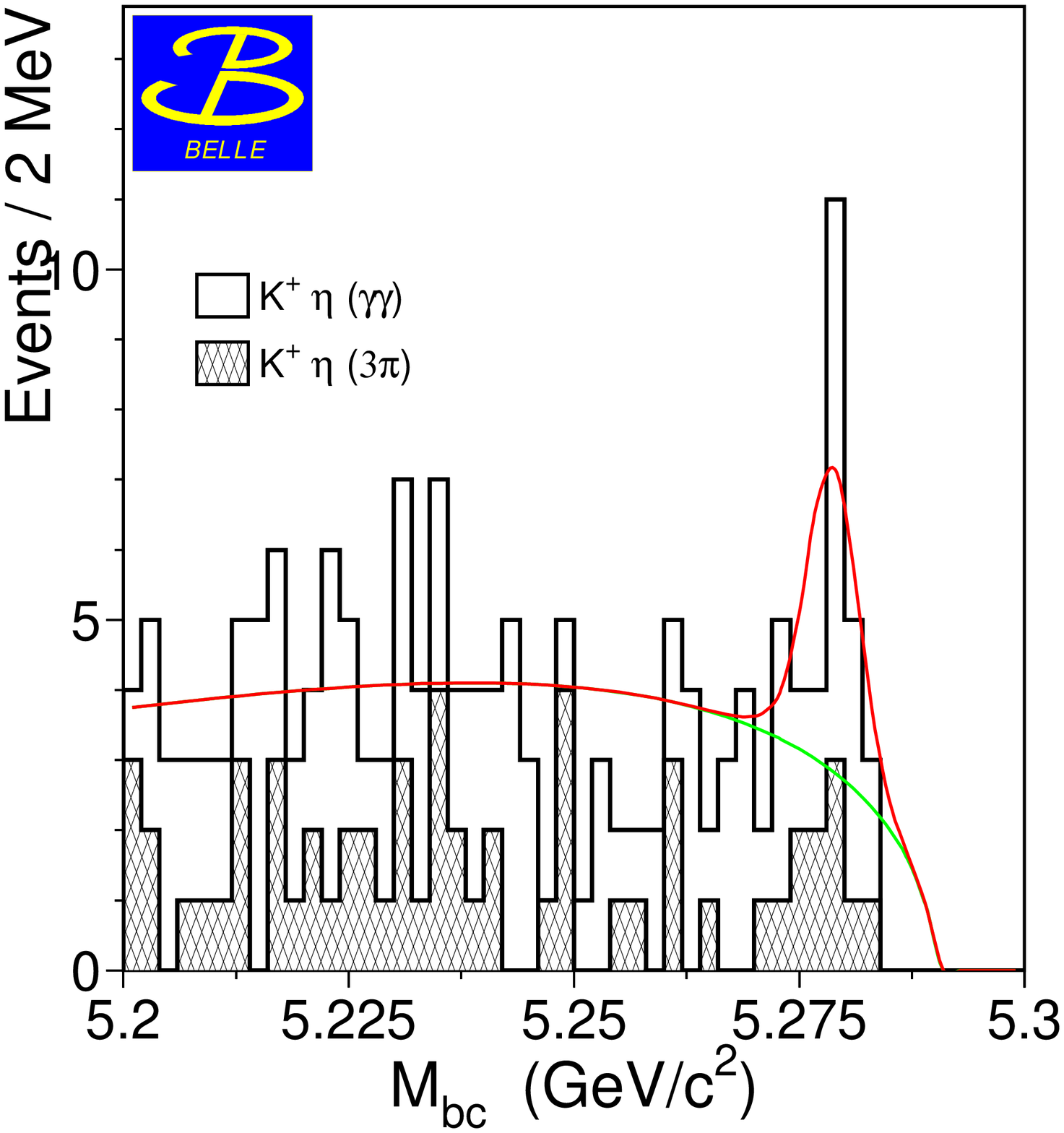}
\includegraphics[width=1.4in]{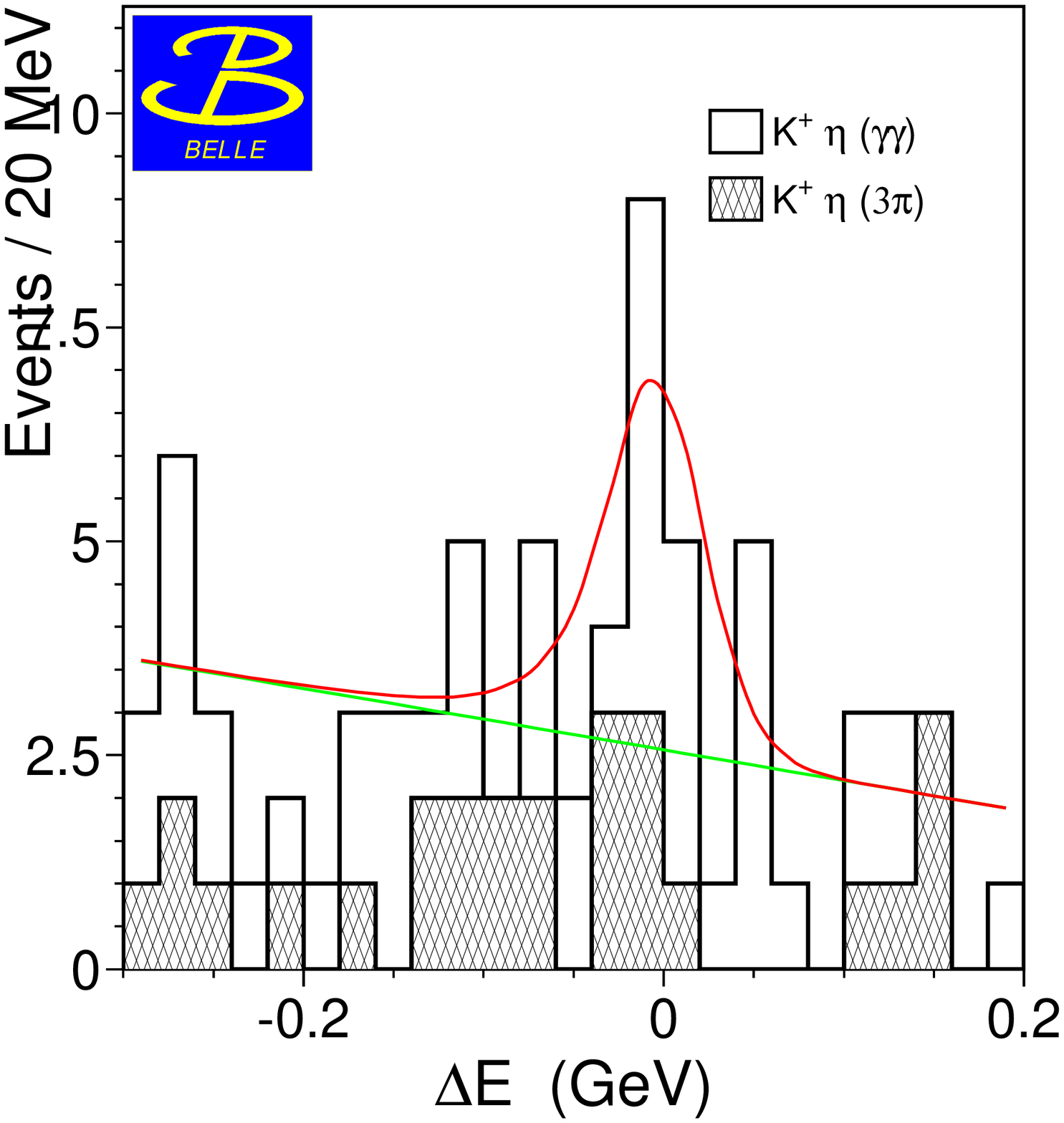} ~~~
\includegraphics[width=1.4in]{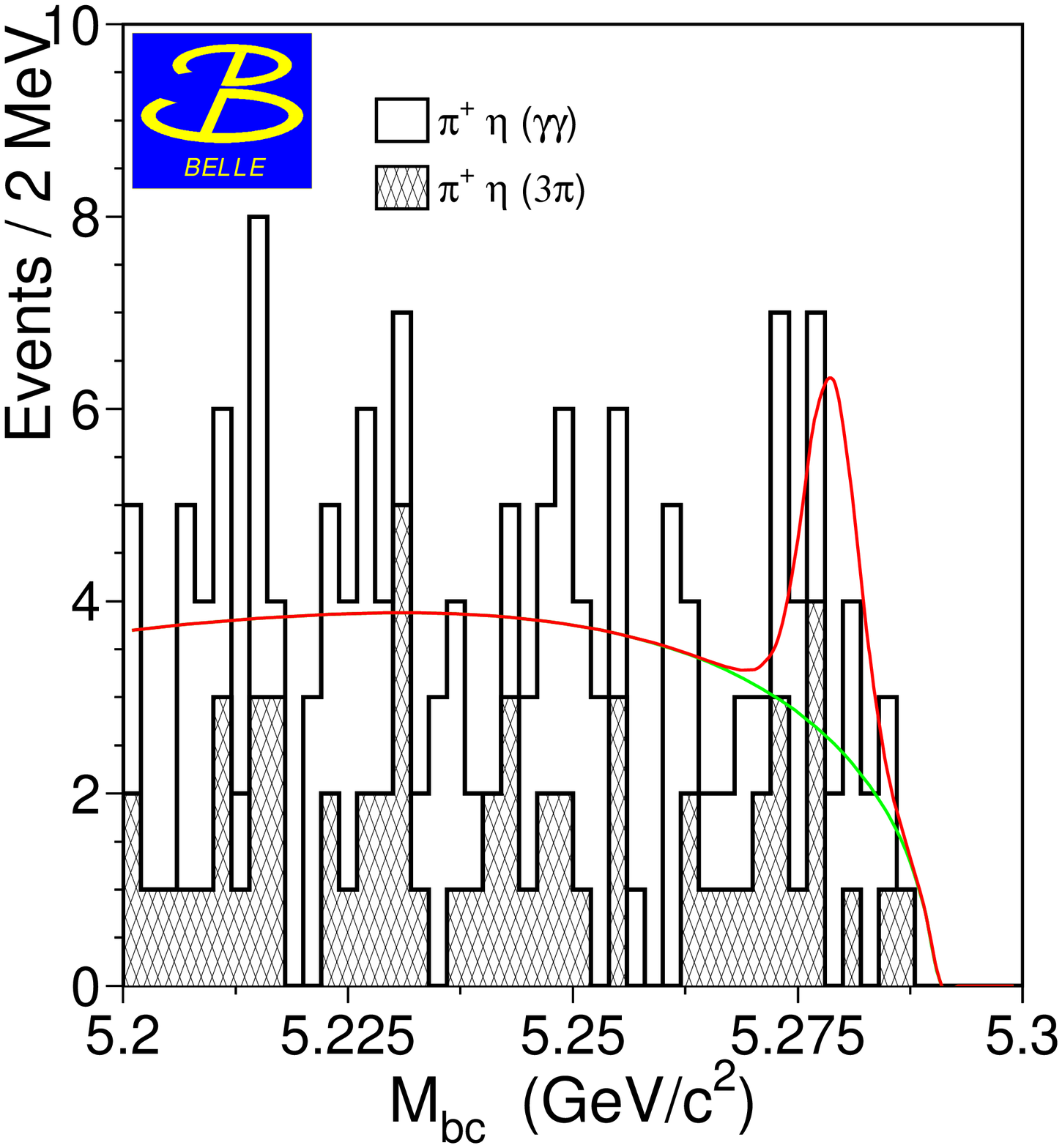}
\includegraphics[width=1.4in]{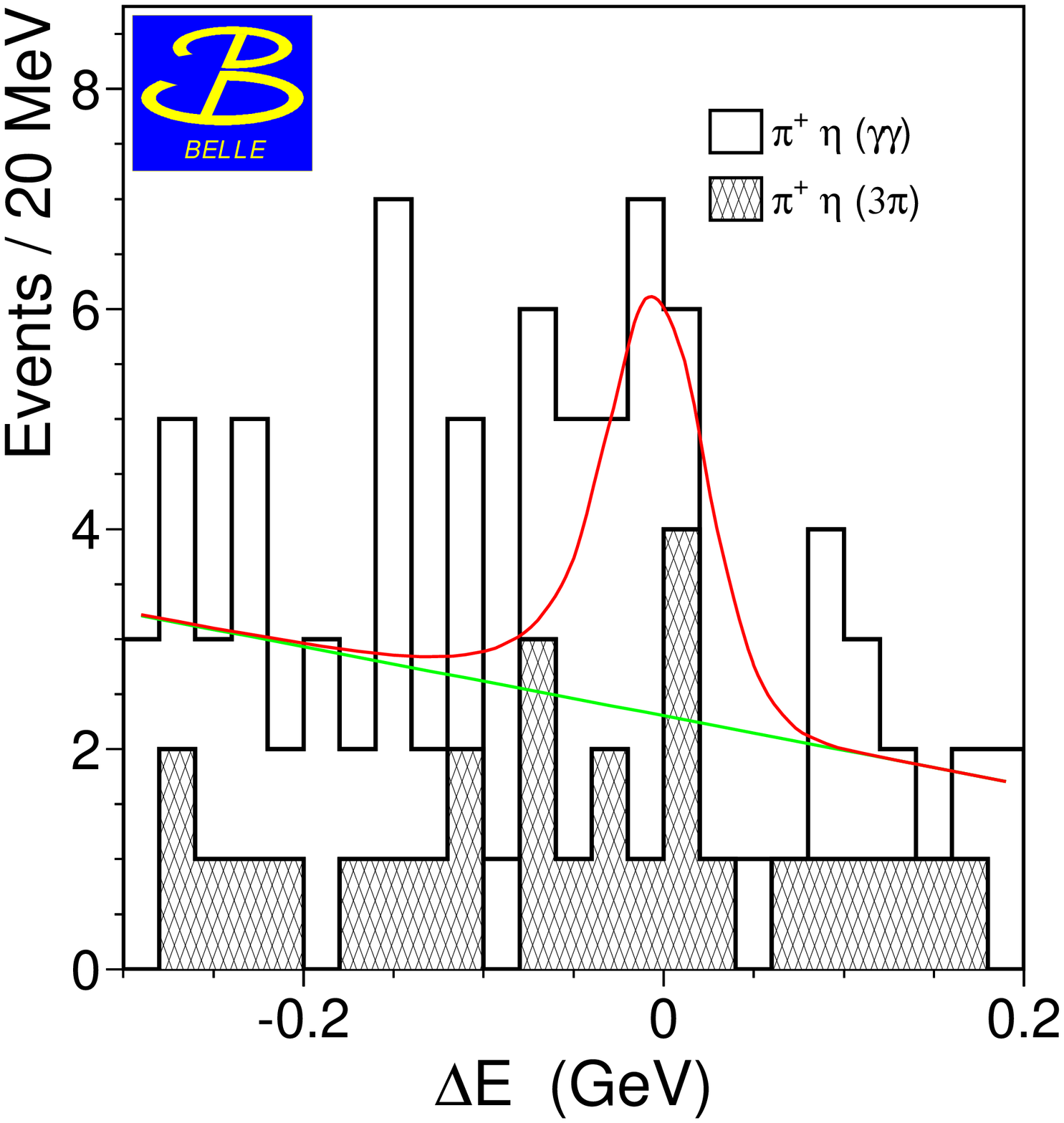}
\end{center}
\caption{$\mb$ and $\Delta E$ distributions for (left) $\btoetak$ and
(right) $\btoetapi$.  Histograms represent data, with the $\etapi$
subset shaded, the solid curves represent the fit functions.
\label{fg:etah}}
\end{figure}

Signal yields are obtained from extended unbinned maximum
likelihood (ML) fits on the variables $\mb$ and $\Delta E$.  The
signal probability density functions (PDF) are Gaussian for $\mb$ and
an empirically determined 
parameterization~\cite{cbline} for $\Delta E$. 
The background PDF are taken to be an empirical 
function~\cite{argus} for $\mb$ 
and a first-order polynomial for $\Delta E$.
The statistical significance is defined as 
$\sqrt{-2{\rm ln}({\cal L}(0)/{\cal L}_{\rm max})}$
where ${\cal L}_{\rm max}$ is the likelihood at the nominal signal
yield and ${\cal L}(0)$ is the likelihood with the signal yield fixed
to zero. 
The results of the fits for yields are given in Table~\ref{tb:2body}.  
Figure~\ref{fg:etah} shows distributions of $\mb$ and $\Delta E$ with
the projections of the fit function.  We see strong evidence for the
decays $\btoetak$ and $\btoetapi$ with statistical significance of
$4.9 \sigma$ and $4.3 \sigma$, respectively.  
Assuming the number of $B^0\overline{B}{}^0$ and $B^+B^-$ pairs to be
equal, we find their branching fractions 
$\mathcal{B}(\btoetak) = ( 5.3 ^{+1.8}_{-1.5} \pm 0.6 ) \times
10^{-6}$ and 
$\mathcal{B}(\btoetapi) = ( 5.4 ^{+2.0}_{-1.7} \pm 0.6 ) \times
10^{-6}$.

%%%%%%%%%%%%%%%%%%%%%%%%%%%%%%%%%%%%%%%%%%%%%%%%%%%%%%%%%%%%%%%%%%%%%%
\subsection{$B^\pm \to \omega h^\pm$}
%%%%%%%%%%%%%%%%%%%%%%%%%%%%%%%%%%%%%%%%%%%%%%%%%%%%%%%%%%%%%%%%%%%%%%

%The decays of $B^-\to \omega K^-$ and $\omega\pi^-$ have a historical
%reason to study since it was the first charmless two body mode found
%in the $B$ decays. 
%Certainly the measurements of the branching fractions of 
%$B^-\to \omega K^-$ and $\omega\pi^-$ decays also provide a chance
%to examine the models of $B$ decay.

In this analysis, candidate $\omega$ mesons are reconstructed from
$\pi^+\pi^-\pi^0$ 
combinations where the CM momentum of the $\pi^0$ is required to be 
greater than 350 MeV/$c$ to reduce the large 
combinatorial background from low energy photons.
The invariant mass of the $\pi^+\pi^-\pi^0$ combination is required to be
within $\pm 30$ MeV/$c^2$ of the nominal $\omega$ mass~\cite{pdg}
(the natural width of the $\omega$ is $8.9$ MeV/$c^2$).

\begin{figure} %[t]
\begin{center}
\includegraphics[width=1.5in]{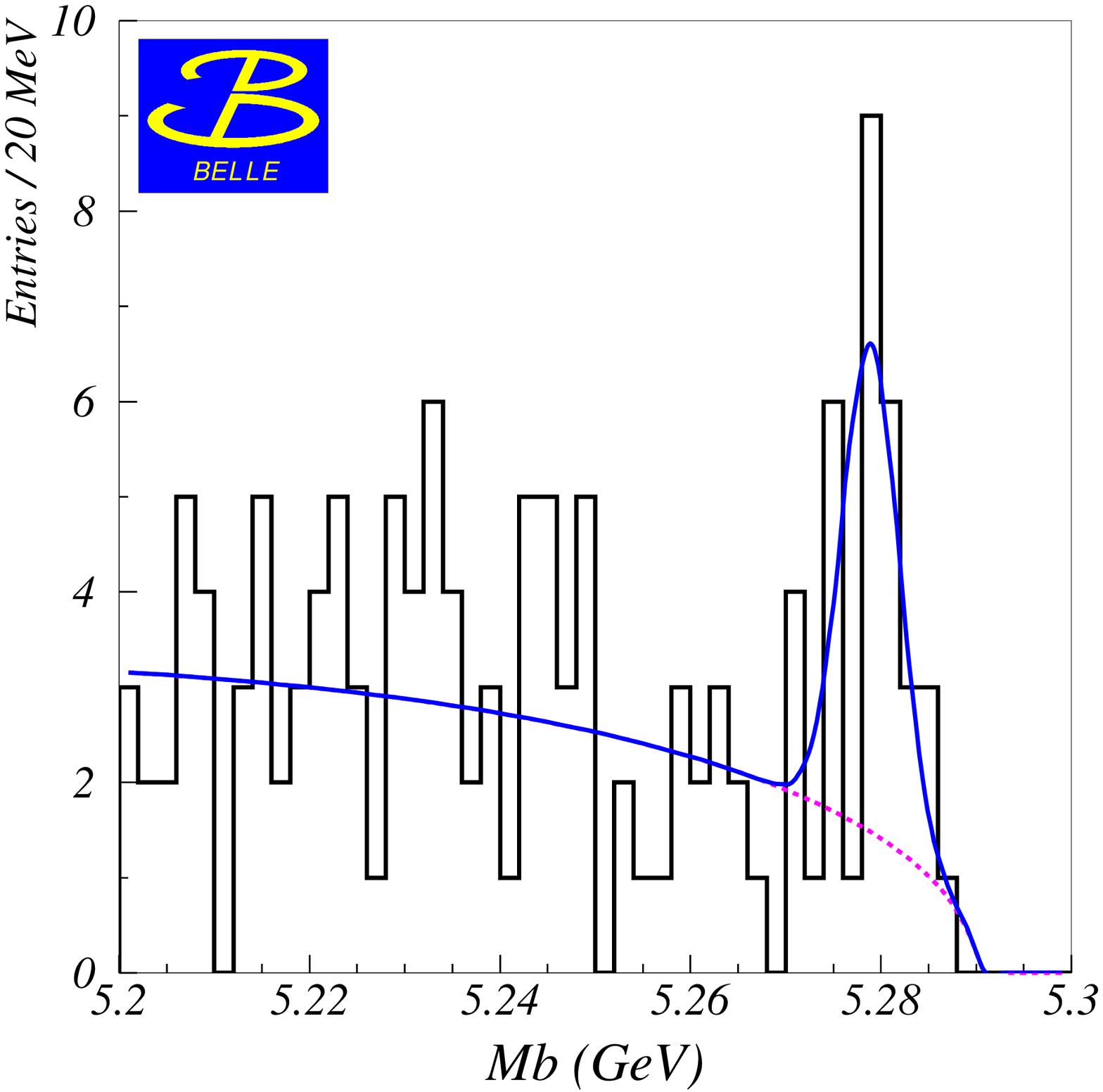}
\includegraphics[width=1.5in]{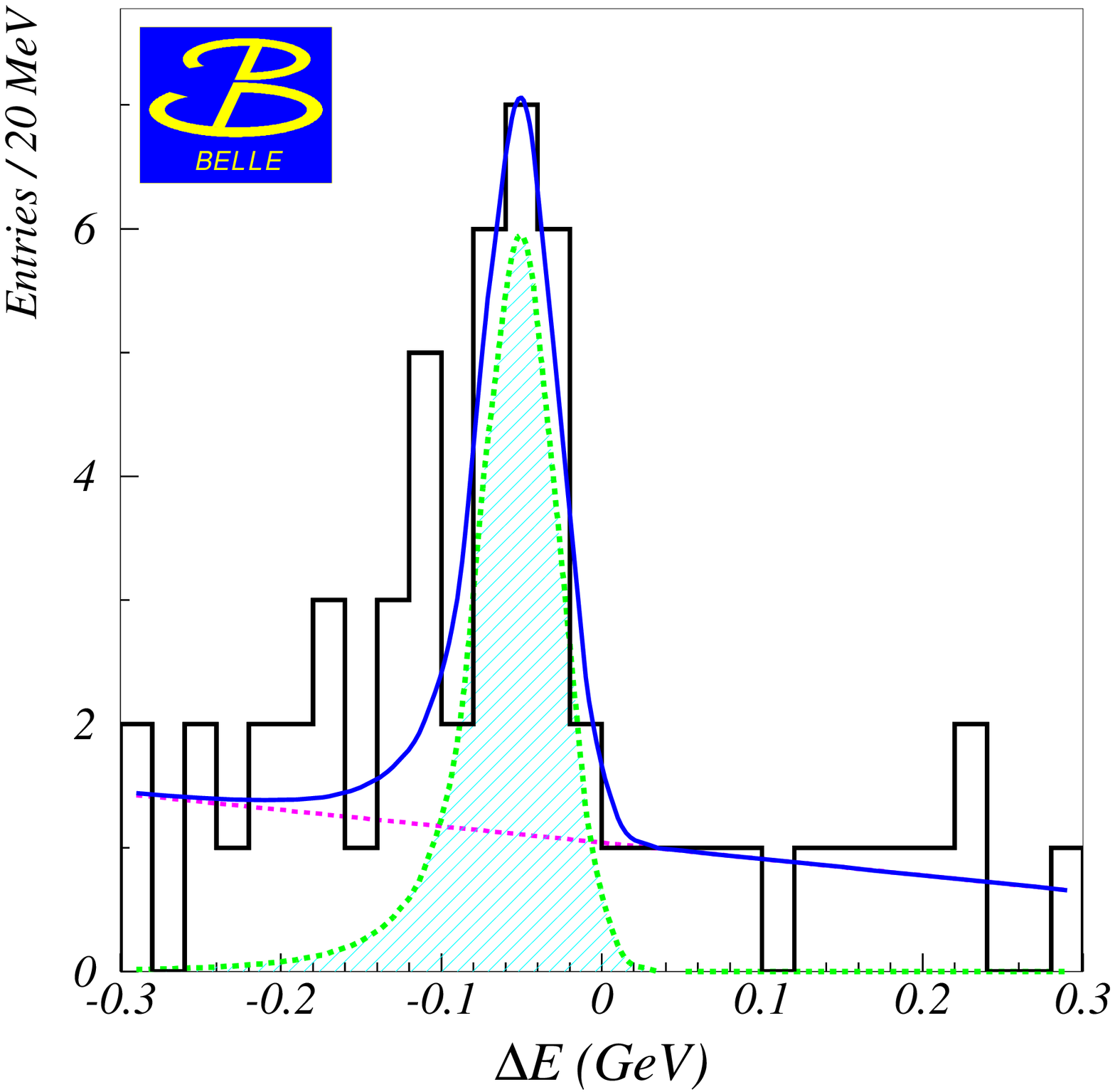}
\includegraphics[width=1.5in]{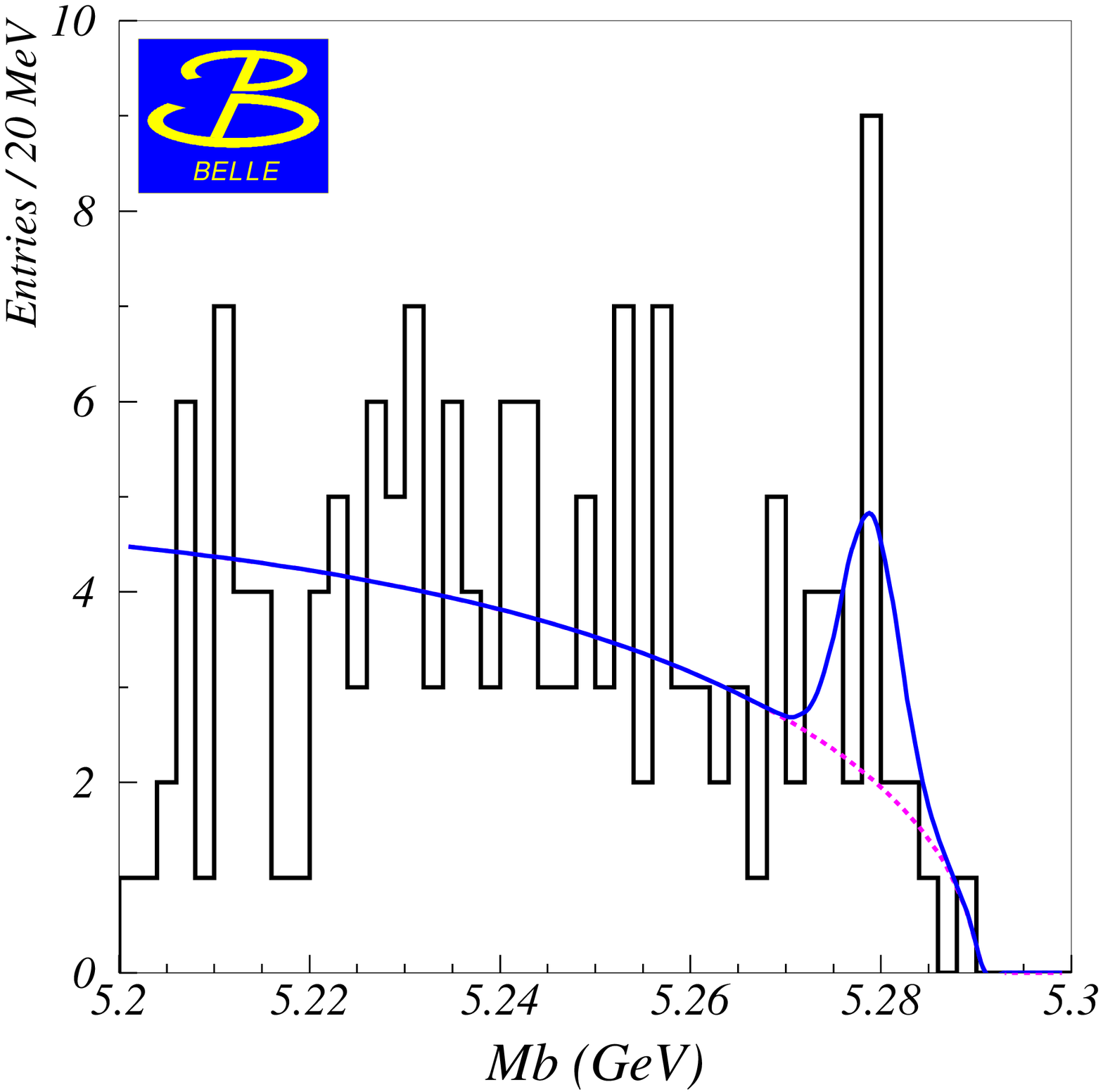}
\includegraphics[width=1.5in]{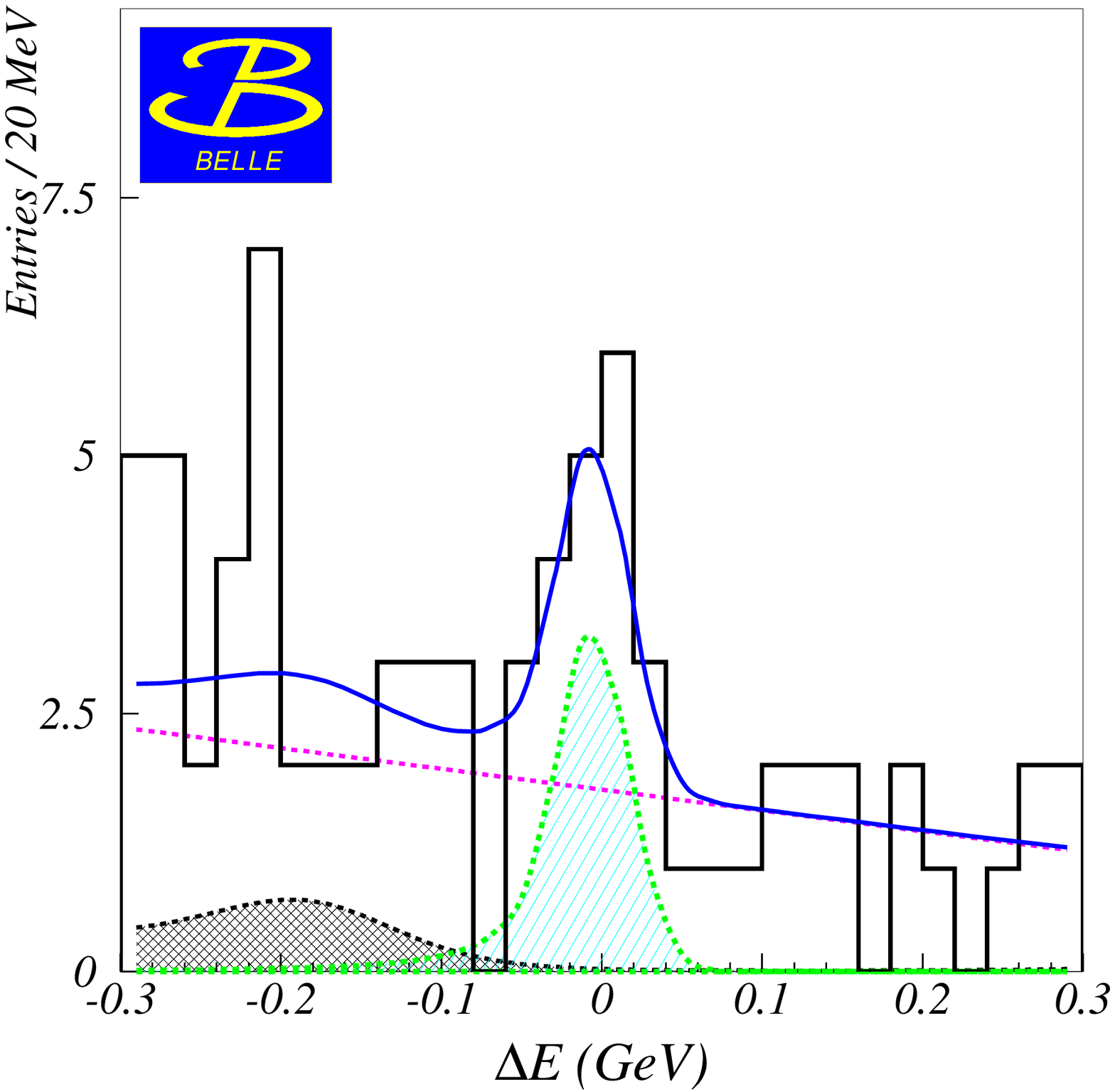}
\end{center}
\caption{The $M_{bc}$ and $\Delta E$ projections with the normalized 
signal and background PDFs for (left) $B^\pm \to \omega K^\pm$ and (right)
$B^\pm \to \omega\pi^\pm$.
\label{omegah}}
\end{figure}

Signal yields are extracted using unbinned ML fit 
simultaneously for $\mb$ and $\Delta E$.
The background functions include a combinatorial component and a
component from other charmless $B$ decays.
Due to the possible mis-identification between $K^-$ and $\pi^-$,
we also include a component for feed-down from $\omega \pi^-$ to 
$\omega K^-$ fitting and vice versa.
If a pion is mis-identified to a kaon, the wrong mass assignment 
shifts the $\omega\pi^-$ signal 44~MeV away from zero, and 
the feed-down component from $\omega\pi^-$ can be distinguished
in the $\Delta E$ distribution.
The results of the fit are 
summarized in the Table~\ref{tb:2body}.
The projections of the $M_{bc}$ and $\Delta E$ with the normalized 
signal and background PDFs are shown in Figure~\ref{omegah}.
We have observed the decay $B^\pm \to \omega K^\pm$ with $6.4\sigma$ 
significance.  Our results on the $B^-\to\omega K^-$ and $\omega\pi^-$ 
branching fraction measurements disagree with the earlier ones from
CLEO~\cite{cleo-prl-85} and BABAR~\cite{babaretapk}.
But the sum of the branching fractions of $B^-\to (\omega K^- +
\omega\pi^-)$ is consistent with the CLEO's result.

%%%%%%%%%%%%%%%%%%%%%%%%%%%%%%%%%%%%%%%%%%%%%%%%%%%%%%%%%%%%%%%%%%%%%%
\section{Three-Body Charmless \boldmath Hadronic $B$ Decays}
%%%%%%%%%%%%%%%%%%%%%%%%%%%%%%%%%%%%%%%%%%%%%%%%%%%%%%%%%%%%%%%%%%%%%%

Belle has recently published results on tree body charmless hadronic
decays $B^\pm \to K^\pm h^+ h^-$\cite{khhprd}.  It is of interest to look
for similar phenomena in the neutral channel, to further investigate
the $b \to s$ penguin transitions which mediate these decays.  In
addition, these modes may in future be used to measure $CP$
violation.

%%%%%%%%%%%%%%%%%%%%%%%%%%%%%%%%%%%%%%%%%%%%%%%%%%%%%%%%%%%%%%%%%%%%%%
\subsection{\boldmath $B^0 \to K^0 h^+ h^-$}
%%%%%%%%%%%%%%%%%%%%%%%%%%%%%%%%%%%%%%%%%%%%%%%%%%%%%%%%%%%%%%%%%%%%%%

\begin{figure}  %[!htb]
\begin{center}
\includegraphics[width=2.2in]{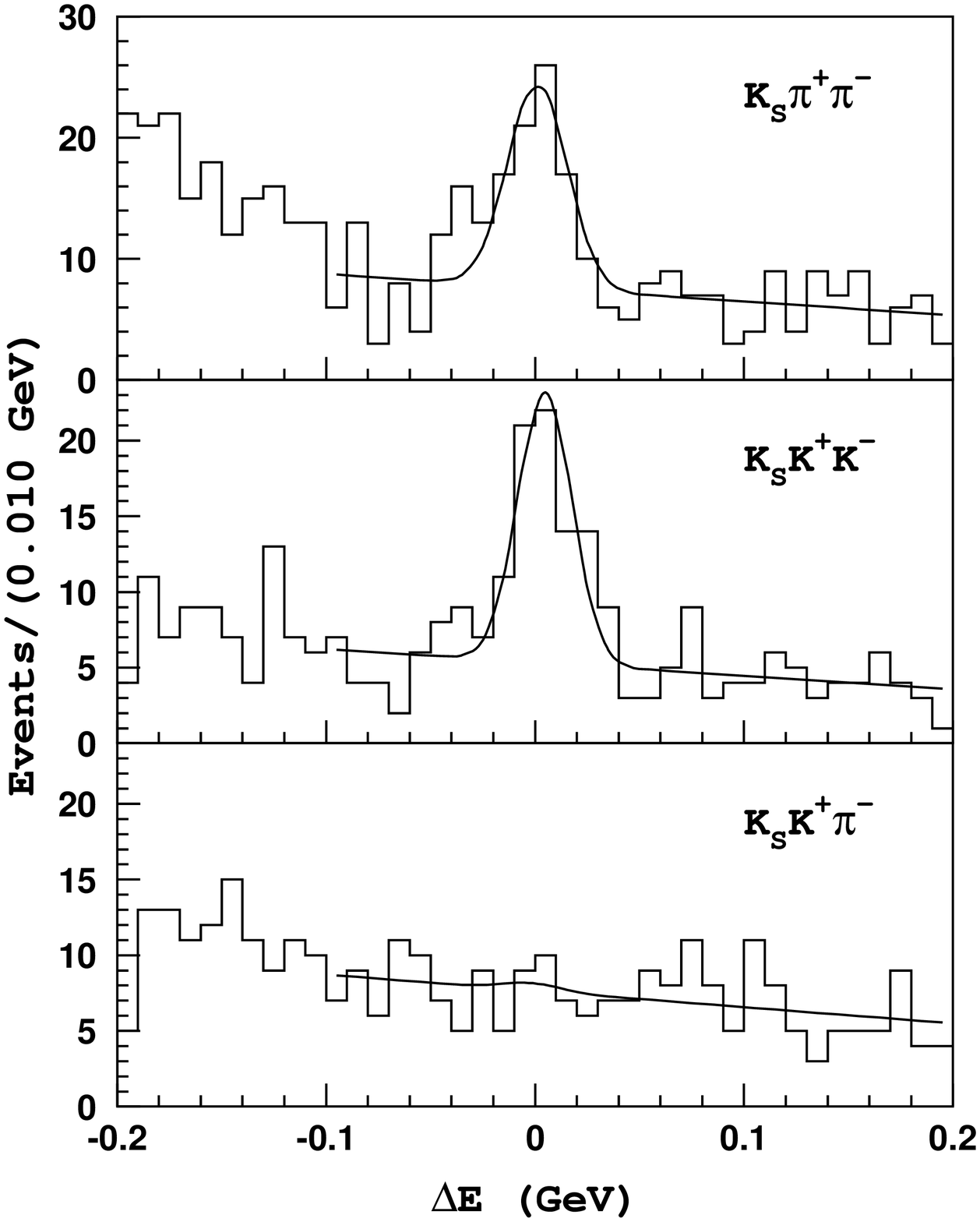}
\includegraphics[width=2.2in]{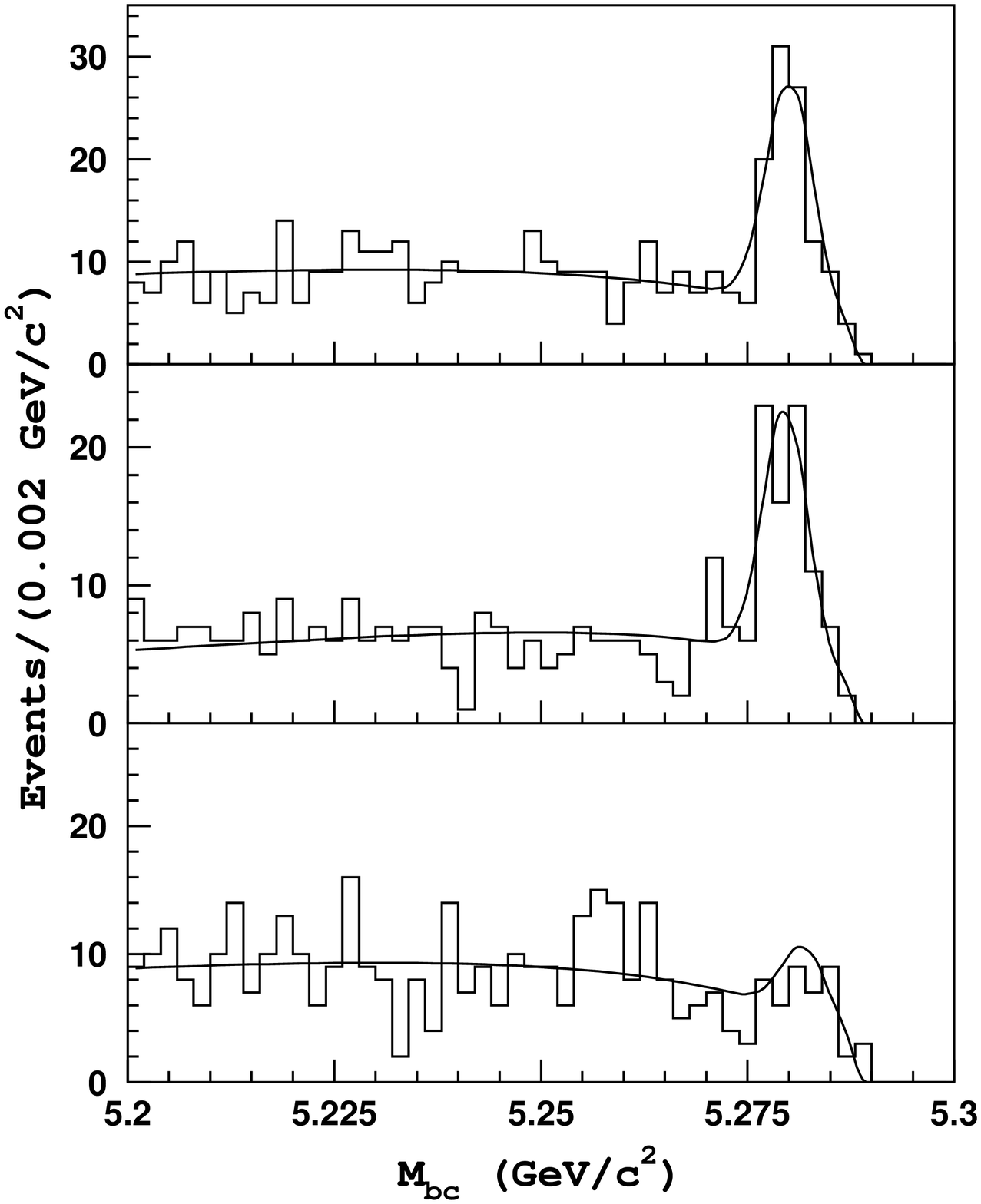}
\end{center}
\caption{The $\Delta{E}$ (left) and $M_{\rm bc}$ (right) distributions 
         for three-body final states: $K_S\pi^+\pi^-$ (top), $K_SK^+K^-$ 
         (middle) and $K_SK^{\pm}\pi^{\mp}$ (bottom).
\label{fg:kshh-mbde}}
\end{figure}

In this analysis, we reconstruct the decays $B^0 \to K^0 h^+ h^-$
without any assumption on the intermediate hadronic resonance.  Only
$K^0 \to K^0_S \to \pi^+ \pi^-$ is considered here.  As in other rare
$B$ decay 
modes, continuum events are the dominant background source and are
suppressed using various event shape and kinematic variables.  In the
$B^0 \to K^0 \pi^+ \pi^-$ mode, we also have 
large backgrounds from $B^0{\to}D^-\pi^+, D^-{\to}K_S\pi^-$ and
$B^0{\to}J/\psi K_S, J/\psi{\to}\mu^+\mu^-$ where muons are
misidentified as pions.
These backgrounds are suppressed by requiring 
$ |M(K_S\pi^-)-M_{D^-}|    > 0.100~\mbox{GeV/}c^2$, 
$ |M(h^+h^-)-M_{J/\psi}|   >0.070~\mbox{GeV/}c^2$, and 
$ |M(h^+h^-)-M_{\psi(2S)}| >0.050~\mbox{GeV/}c^2$, 
where $h^+$ and $h^-$ are pion candidates. 
Signal yields are obtained from fits to the $\Delta E$ distributions.
We find $60.3 \pm 11.0$ $B^0 \to K^0 \pi^+ \pi^-$ events and $57.9 \pm
10.0$ $B^0 \to K^0 K^+ K^-$ events, 
which corresponds to preliminary branching fractions:
\begin{eqnarray*}
{\mathcal{B}}(B^0{\to}K^0\pi^+\pi^-) &=&(\brkspp) \times 10^{-6}, 
	\,\,\,(6.6\sigma) \\
{\mathcal{B}}(B^0{\to}K^0  K^+  K^-) &=&(\brkskk) \times 10^{-6}
	\,\,\,(7.4\sigma) 
\end{eqnarray*}
Figure~\ref{fg:kshh-mbde} shows the $\Delta E$ and $\mb$
distributions for these decays after the background suppression.  We
also search for $B^0 \to K^0 K^\pm \pi^\mp$ but do not observe significant
signals.  The $90\%$ C.L. braching fraction upper limit is calculated
to be $\mathcal{B}(B^0 \to K^0K^{\pm}\pi^{\mp}) < \brkskp \times 10^{-6}$.

\begin{figure} [b]
\begin{center}
\includegraphics[width=1.5in]{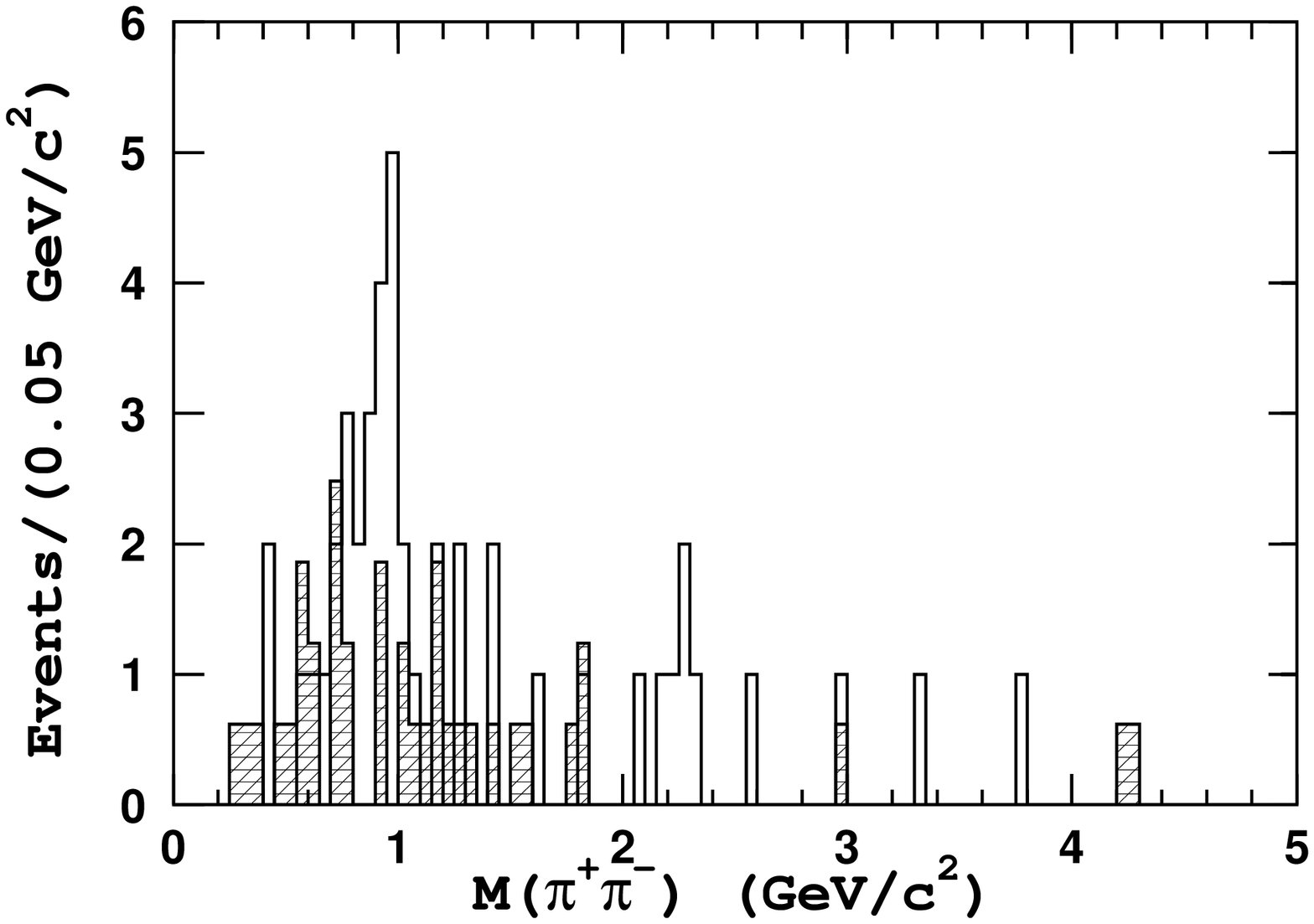}
\includegraphics[width=1.5in]{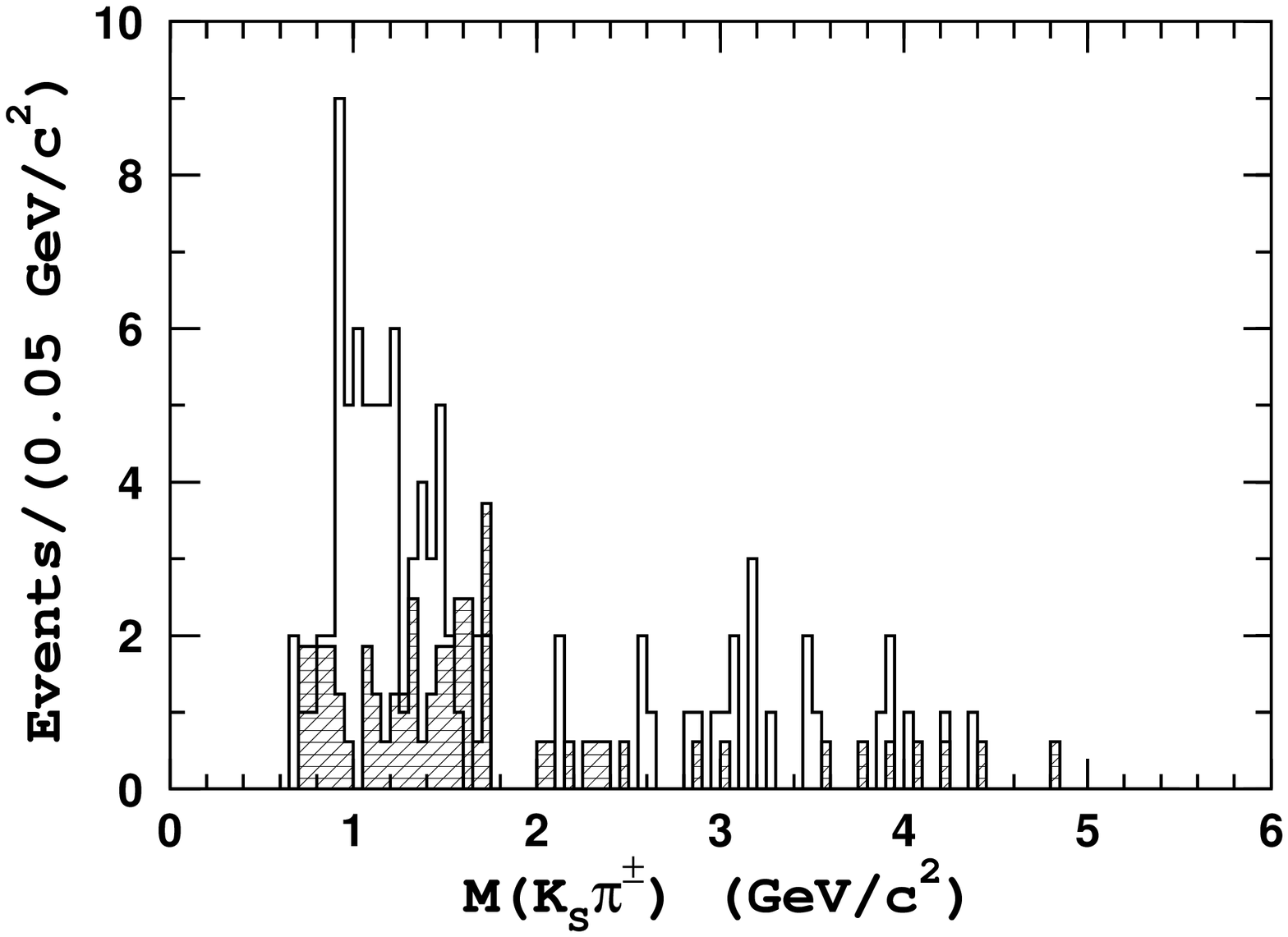}
\includegraphics[width=1.5in]{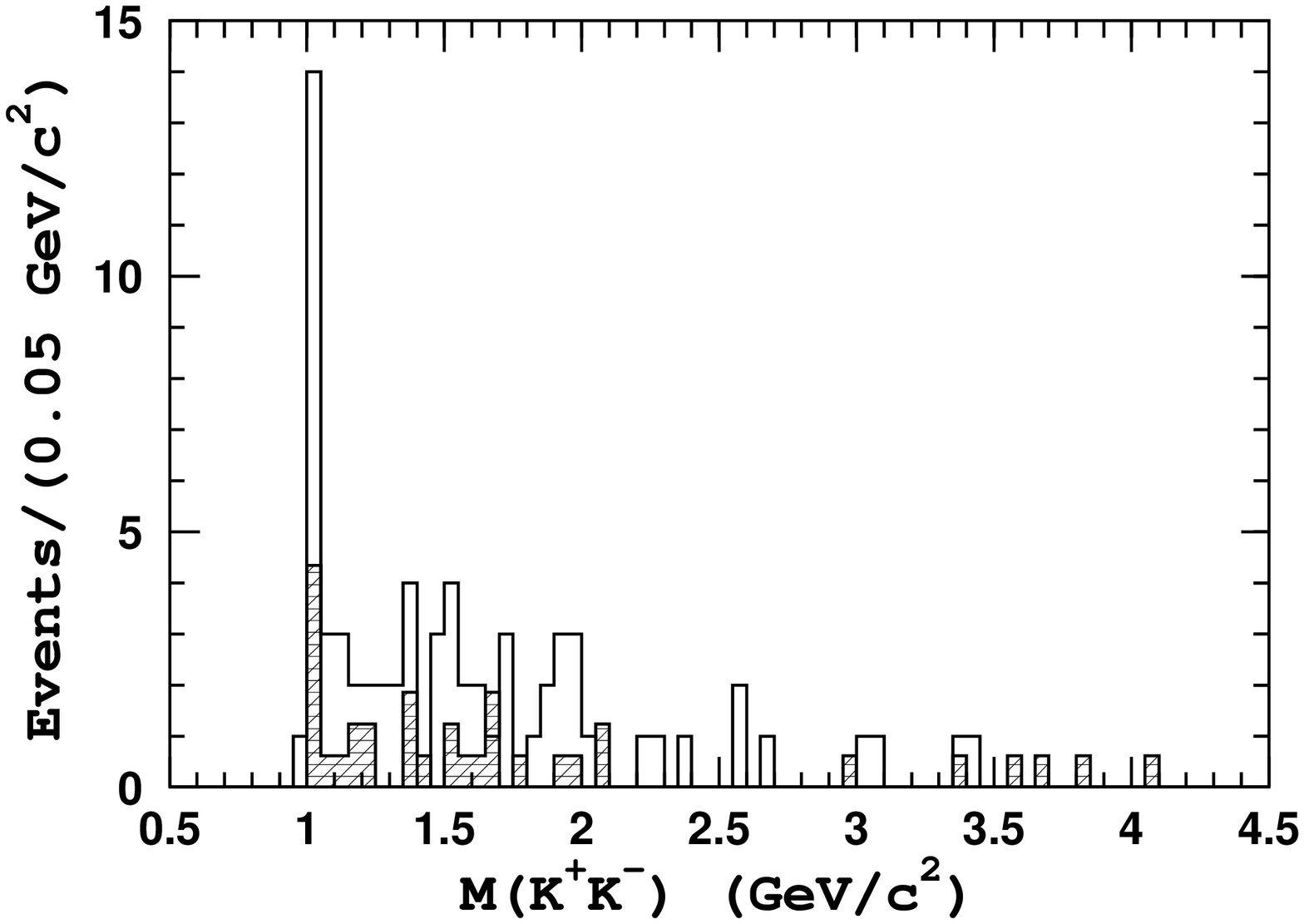}
\includegraphics[width=1.5in]{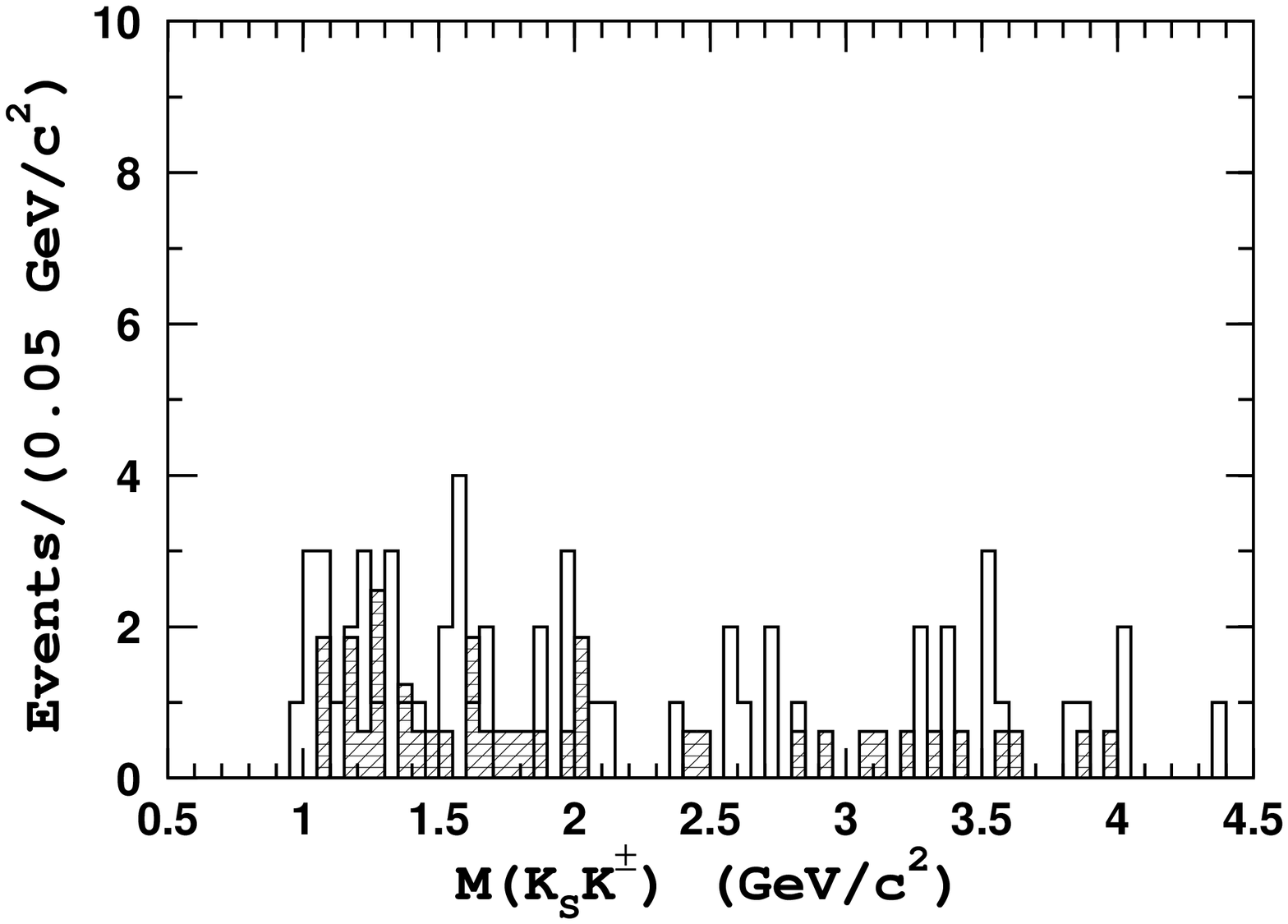}
\end{center}
\caption{(a) $\pi^+ \pi^-$ and (b) $\pi^+ K^0_S$ invariant mass
spectra for selected $B^0 \to K^0_S \pi^+ \pi^-$, (c) $K^+ K^-$ and
(d) $K^+ K^0_S$ for $B^0 \to K^0_S K^+ K^-$, signal events 
(open histograms) and for background events in the $\Delta E$
sidebands (hatched).
\label{fg:kshh-mhh}}
\end{figure}

Further studies of intermediate resonant states of these decays are
made with a Dalitz plot style analysis.  Figure~\ref{fg:kshh-mhh} shows
the $\pi^+ \pi^-$ and $K^0_S \pi^\pm$ invariant mass distributions for
selected $B^0 \to K^0_S \pi^+ \pi^-$ candidates in the $B$ signal
region, and the $K^+ K^-$ and $K^0_S K^\pm$ invariant mass for 
$B^0 \to K^0_S K^+ K^-$.  Clear contributions from $K^*(892)^\pm
\pi^\mp$ 
and $\phi(1020) K^0$ are seen.  We also find broad resonances in 
$K^0_S \pi^\pm$ and $K^+ K^-$ mass around 1.4 GeV$/c^2$ and 1.5 
GeV$/c^2$, respectively.  Assuming a set of two-body final states, we 
perform a simultaneous likelihood fit to the $\Delta E$ distributions 
for various resonance regions and determine the exclusive branching 
fractions 
%neglecting the effects of interference.  
The uncertainty due
to possible interference between different intermediate states is
included as a model-dependent error.  The results are summarized in
Table~\ref{tb:kshh-2body}.

\begin{table}  %[!htb]
\caption{Preliminary results of the simultaneous fits to the 
$B^0 \to K^0\pi^+\pi^-$ and $K^0K^+K^-$ final states, 
respectively.  Branching fractions of the corresponding decay modes
from $B^+ \to K^+ h^+ h^-$ are also listed for 
comparison.~\protect\cite{khhprd} 
\label{tb:kshh-2body}}
\vspace{0.4cm}
\small
\begin{center}
\begin{tabular}{|l|ccc|c|c|} \hline
\begin{tabular}{c}
Mode\\
\end{tabular} &
\begin{tabular}{c}
$\epsilon$ (\%) \\
\end{tabular} &
\begin{tabular}{c}
Yield \\
\end{tabular} &
\begin{tabular}{c}
$\Sigma$\\
\end{tabular} &
\begin{tabular}{c}
${\cal{B}}(B^0{\to}Rh){\times}$\\${\cal{B}}(R{\to}hh)$ ($\times 10^{-6}$)\\
\end{tabular} &
\begin{tabular}{c}
${\cal{B}}(B^+{\to}Rh^+){\times}$\\${\cal{B}}(R{\to}h^+h^-)$~\cite{khhprd} 
($\times 10^{-6}$) \\
\end{tabular} \\
\hline
$K^*(892)^{\pm}\pi^{\mp}$  & $4.87\pm0.23$ & $16.9^{+6.0}_{-5.2}$ & 3.9
& $\brksppkstpi$ & $\brkcppkstpi$ \\
$K_X(1400)^{\pm}\pi^{\mp}$ & $4.22\pm0.21$ & $24.9^{+9.0}_{-8.3}$ & 3.3
& $\brksppkstxpi$ & $\brkcppkstxpi$ \\ 
$\rho^0(770) K^0$          & $4.56\pm0.23$ & 
$~1.4^{+6.4}_{-5.7}$ & 
0.2 & $<\brkspprhoks$ & $<\brkcpprhokc~(90\%CL)$ \\
$f_0(980) K^0$             & $5.22\pm0.24$ & 
$~9.4^{+6.0}_{-4.9}$ & 
2.1 & $<\brksppfzeroks$ & $\brkcppfzerokc$ \\
$f_X(1300) K^0$            & $5.27\pm0.24$ & 
$~8.0^{+6.0}_{-5.0}$ &
1.7 & $<\brksppfxks$ & $\brkcppfxkc$ \\
\hline
$\phi(1020) K^0$ & $7.01\pm0.19$ & $11.7_{-4.6}^{+5.5}$ & 3.0 & $\brkskkphiks$ & $\brkckkphikc$ \\ 
$f_X(1500)K^0$   & $6.25\pm0.18$ & $33.5_{-7.5}^{+8.1}$ & 5.3 & $\brkskkfxks$ & $\brkckkfxkc$ \\ 
$a_0(980)^{\pm}K^{\mp}$  & $2.70\pm0.38$ & 
$3.4^{+3.0}_{-2.4}$ &
1.5 & $<\brkskkak$ & --- \\
$a_X(1300)^{\pm}K^{\mp}$ & $4.45\pm0.09$ & 
$4.0^{+4.7}_{-4.1}$ &
1.0 & $<\brkskkaxk$ & --- \\
\hline
\end{tabular} 
\end{center}
\end{table}

%%%%%%%%%%%%%%%%%%%%%%%%%%%%%%%%%%%%%%%%%%%%%%%%%%%%%%%%%%%%%%%%%%%%%%
\section{Charmless \boldmath Baryonic $B$ Decays}
%%%%%%%%%%%%%%%%%%%%%%%%%%%%%%%%%%%%%%%%%%%%%%%%%%%%%%%%%%%%%%%%%%%%%%

In contrast to charm meson decay, final states
with baryons are allowed in $B$ meson decay. 
To date, a few low multiplicity
$B$ decay modes with baryons in the final state from
$b\to c$ transitions have been observed.~\cite{cleobcbary}  
Rare $B$ decays due to charmless
$b\to s$ and $b\to u$ transitions should also lead to final states
with baryons. A number of searches for such modes have been carried out
by CLEO,~\cite{cleorrbary} ARGUS,~\cite{argusbary} 
and LEP~\cite{delphibary} but only upper limits were obtained. 
Stringent upper limits for two-body modes such as $B \to p \bar{p}$, 
$\bar{\Lambda} p$, and $\Lambda \bar{\Lambda}$ have recently been 
reported by Belle.~\cite{bellebary}

%%%%%%%%%%%%%%%%%%%%%%%%%%%%%%%%%%%%%%%%%%%%%%%%%%%%%%%%%%%%%%%%%%%%%%
\subsection{$B^\pm \to p \overline{p} K^\pm$ \protect\cite{ppk}}
%%%%%%%%%%%%%%%%%%%%%%%%%%%%%%%%%%%%%%%%%%%%%%%%%%%%%%%%%%%%%%%%%%%%%%

We have searched for the decay modes 
$B^\pm \to  p \bar{p} K^\pm$ and $B^0\to p \bar{p} K^0_S$.
These modes are expected to proceed mainly via $b\to s$ penguin
diagrams. %~\cite{Kpi}  
We also search for
$B^{+}\to p \bar{p} \pi^{+}$ which is expected
to occur primarily via a $b\to u$ tree process. Once they are
established, these baryonic modes 
may be used to either constrain or observe direct $CP$ violation 
in $B$ decay.~\cite{rosner_baryon}

To reconstruct signal candidates in the $B^+\to  p \bar{p} K^+$ mode, 
we form combinations of a kaon, proton and anti-proton
that are inconsistent with the following $b\to c \bar{c} s$ transitions:
$B^+\to J/\psi K^+$, $J/\psi \to p \bar{p}$; $B^+\to \eta_c K^+$,
$\eta_c\to p \bar{p}$; $B^+\to \psi^{\prime} K^+$, 
$\psi^{\prime}\to p \bar{p}$
and $B^+\to \chi_{c[0,1]} K^+$, $\chi_{c[0,1]} \to p \bar{p}$.
This set of requirements is referred to as the charm veto. 
Similar charm vetoes are applied in the analysis of
the other decay modes. In the case of $B^0\to p \bar{p} K_S$,
events with $p K_S$ or $\bar{p} K_S$ 
masses consistent with the $\Lambda_c$ are rejected. 

\begin{figure}
\begin{center}
\includegraphics[height=1.8in]{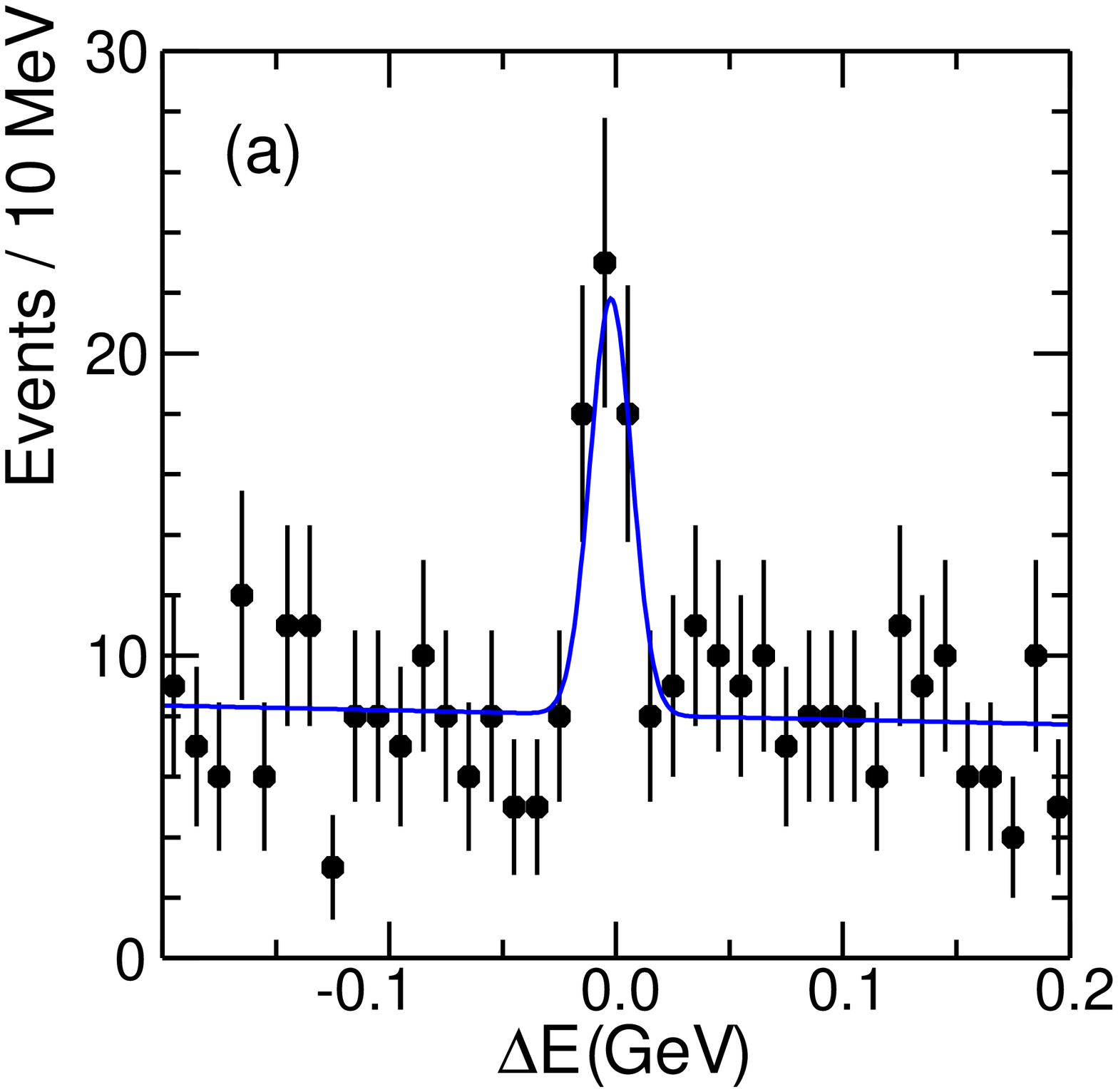}
\includegraphics[height=1.8in]{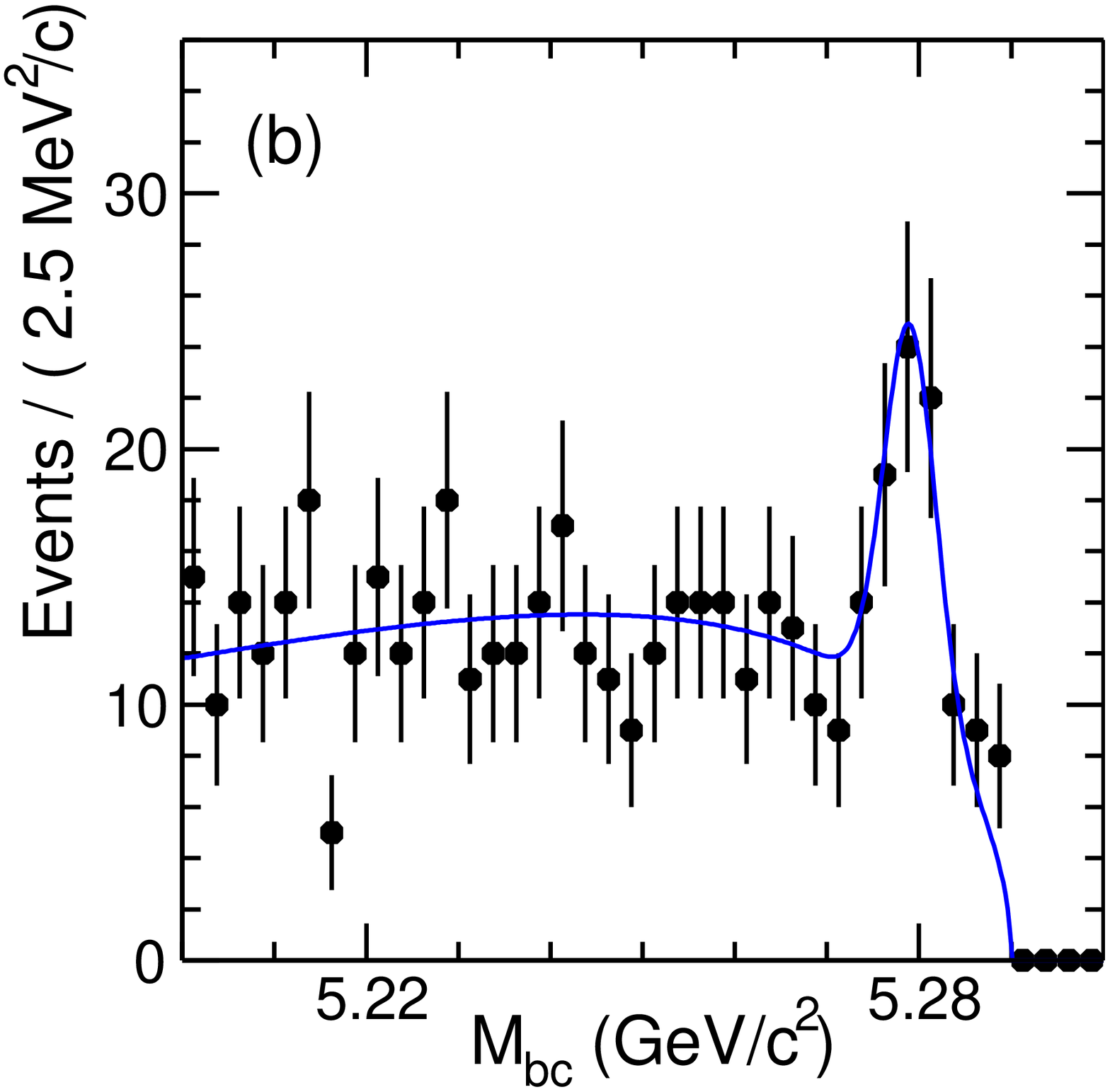}
\includegraphics[height=1.8in]{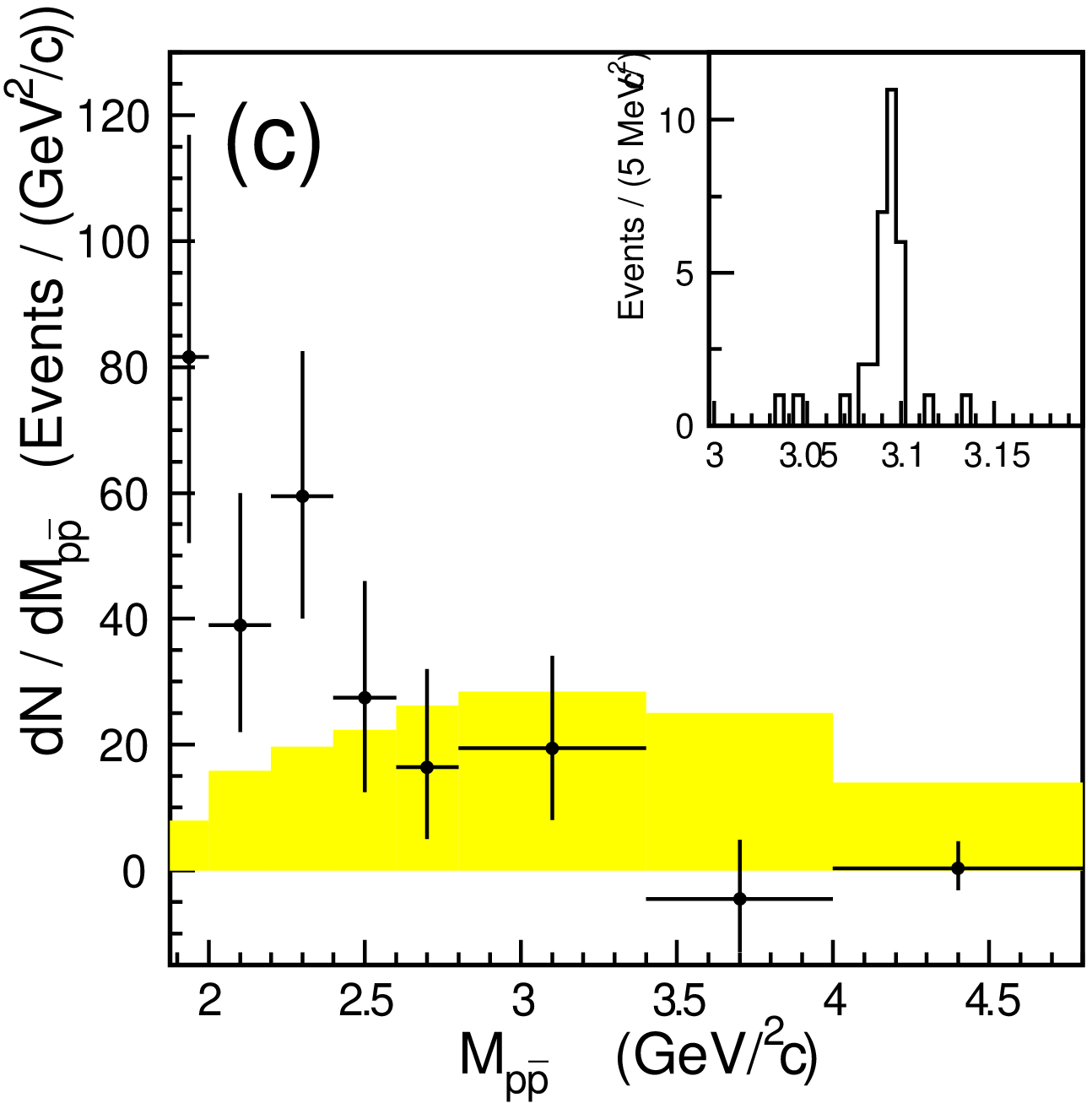}
\end{center}
\caption{(a) $\Delta E$ and (b) $\mb$
distributions for         
$B^+\to p\bar{p} K^+$ candidates.
(c) The fitted yield divided by the bin size for 
$B^+\to p\bar{p} K^+$ as a function of $M_{p \bar{p}}$.
%The charm veto is applied. 
The distribution from non-resonant $B^+\to p \bar{p} K^+$ 
MC simulation is superimposed.  The inset shows the $M_{p \bar{p}}$ 
distribution for the $J/\psi K^+$ signal region by removing the charm veto.
\label{mbde_ppk}}
\end{figure}

In Figure~\ref{mbde_ppk}, 
we show the $\Delta E$ distribution (with $5.27 
{\rm ~GeV}/c^2<M_{\rm bc}<5.29$ GeV/$c^2$)  
and $\mb$ distribution (with $|\Delta E| < 50$ MeV) 
for the signal candidates. 
We fit the $\Delta E$
distribution with a double Gaussian for signal 
and a linear background function with slope determined from
the $\mb$ sideband. 
The fit to the $\Delta E$ distribution gives a yield
of $42.8^{+10.8}_{-9.6}$ with a significance of $5.6 \sigma$.
This is the first observation of a $b\to s$ transition with baryons in
the final state.  

We also examine the $M_{p \bar{p}}$ mass distributions for events
in the $\Delta E, ~M_{\rm bc}$ signal region. The signal yield as a function
of $p\bar{p}$ mass is shown in Figure~\ref{mbde_ppk}(c). These yields were
determined by fits to the $\Delta E$ distribution in bins of $p \bar{p}$
invariant mass. The distribution from a three-body phase space 
MC normalized to the area of
the signal is superimposed. It is clear that
the observed mass distribution is not consistent with three-body
phase space but instead is peaked at low $p\bar{p}$ mass.
This feature is suggestive of 
quasi two-body decay.~\cite{glueball}  It is also possible that the
decay is a 
genuine three-body process and that this feature of the $M_{p \bar{p}}$
spectrum is a baryon form factor effect.~\cite{hou_baryon,cheng_baryon}
%
%We also examine the $p K^-$ mass distribution but do not
%observe any obvious narrow structures such as the $\Lambda(1520)$.

To avoid model dependence in the determination of the branching
fraction for $p \bar{p} K^+$, we fit the $\Delta E$ signal yield in 
bins of $M(\_{p \bar{p}}$
and correct for the detection efficiency in each bin
using a three-body phase space
$B^+\to p \bar{p} K^+$ MC model. 
We then sum the partial branching
fractions in each bin to obtain
$${\cal B}(B^+\to p \bar{p} K^+) =(4.3^{+1.1}_{-0.9}({\rm stat})\pm 0.5({\rm
syst}))\times 10^{-6}.$$

To verify the analysis procedure and branching fraction
determination, 
we remove the $J/\psi$ veto and examine the decay chain
$B^+\to J/\psi K^+$, $J/\psi\to p \bar{p}$. 
The $p\bar{p}$ invariant mass spectrum for $J/\psi K^+$ 
signal candidates is shown as an inset in Figure~\ref{mbde_ppk}(c).  
The obtained branching fraction is in good agreement with 
the PDG world average.

For $B^0\to p \bar{p} K^0_S$ and $B^+\to p \bar{p}\pi^+$ modes, after
the application of the charm and $\Lambda_c$ vetoes,
no significant signals are observed.  A fit to the $\Delta E$
distributions to $\ppks$ gives $6.4^{+4.4}_{-3.7}$ events
and $\pppi$ gives $16.2^{+8.6}_{-8.0}$ events with significance of
$2.1\sigma$.  Applying the Feldman-Cousins procedure,~\cite{feldman},
we obtain an upper limit at 90$\%$ C.L. of $\mathcal{B}(B^0\to p \bar{p}
K^0)<7.2 \times 10^{-6}$ and $\mathcal{B}(B^+\to p \bar{p}
\pi^+)<3.7\times 10^{-6}$.

%%%%%%%%%%%%%%%%%%%%%%%%%%%%%%%%%%%%%%%%%%%%%%%%%%%%%%%%%%%
\section{Search for Direct $CP$ Violation}
%%%%%%%%%%%%%%%%%%%%%%%%%%%%%%%%%%%%%%%%%%%%%%%%%%%%%%%%%%%

The most straightforward indication for $CP$ violation in the $B$ meson
system would be a time-independent rate asymmetry between $CP$
conjugate decays into flavor specific or self-tagging final states. 
Direct $CP$ violation (DCPV) of this type will occur
in a decay containing 
at least two amplitudes that have different $CP$ conserving and $CP$
violating phases.  The charge asymmetry will be most evident in decays
where the two amplitudes are of comparable strength.  The partial rate
asymmetry can be written as
$${\cal A}_{\rm CP} = {N(\bar{B}\rightarrow \bar{f}) -
N(B\rightarrow f)  \over N(\bar{B}\rightarrow \bar{f}) +
N(B\rightarrow f) } = {2|A_1||A_2|\sin{\delta}\sin{\phi} \over |A_1|^2
+ |A_2|^2 + 2|A_1||A_2|\cos{\delta}\cos{\phi} }, $$
where $\delta$ and $\phi$ are the $CP$ conserving and $CP$
violating relative phases between amplitudes $A_1$ and $A_2$; $B$
represents either a $B^0_d$ or $B^+$ meson, $f$ represents a self
tagging final state, and $\bar{B}$ and $\bar{f}$ are the conjugate states. 
In the Standard Model, DCPV occurs in charmless hadronic $B$ decay
modes that involve both 
penguin (P) amplitudes and weak $b\rightarrow u$ tree (T) amplitudes
containing the CP violating weak phase $\phi_3 = \arg{
(V^*_{ub})}$ (in a standard convention~\cite{bib:wolfenstein}).

We search for direct $CP$ violation in the $B \to K^+\pi^-$, $K^+
\pi^0$, $K^0_S \pi^+$, and $\omega K^+$ decay modes by measuring the
difference between the yields of $\bar{B}$ and $B$ decays into the
self tagging final states.  
The asymmetries obtained are summarized in Table~\ref{tb:dcpv}. 
All the results are consistent with null asymmetry, hence we set 90\% C.L. 
limits.  Note that the systematic bias from charge asymmetry in the
detectors is less than 1\%, much smaller than the statistical errors
at present.

\begin{table}
\caption{Results of searches for direct $CP$ violation.
\label{tb:dcpv}}
\vspace{0.4cm}
\small
\begin{center}
\begin{tabular}{|c|cc|cc|}
\hline
mode&
N($\bar{B}$)&
N(${B}$)&
$\mathcal{A}_{\rm cp}$&
90\% C.L.
\\
\hline
$K^+\pi^-$&
$103 \pm 12$&
$115 \pm 14$&
$ -0.06 \pm 0.08 \pm 0.01$&
-0.20:0.09
\\
$K^+\pi^0$&
$28 \pm 8$&
$30 \pm 8$&
$ -0.04 \pm 0.19 \pm 0.03$&
-0.39:0.30
\\
$K^0_S\pi^+$&
$49 \pm 8$&
$18 \pm 6$&
$ 0.46 \pm 0.15 \pm 0.02$&
0.18:0.73
\\
$\pi^+\pi^0$&
$24 \pm 8$&
$13 \pm 7$&
$ 0.31 \pm 0.31 \pm 0.05$&
-0.25:0.89
\\
\hline
$\omega K^+$ &
$7.4 \pm 3.5$ &
$11.6 \pm 3.7$ &
$-0.22 \pm 0.27 \pm 0.04$ &
$-0.70 : 0.26$
\\
\hline
\end{tabular}
\end{center}
\end{table}

%%%%%%%%%%%%%%%%%%%%%%%%%%%%%%%%%%%%%%%%%%%%%%%%%%%%%%%%%%%
\section{Summary}
%%%%%%%%%%%%%%%%%%%%%%%%%%%%%%%%%%%%%%%%%%%%%%%%%%%%%%%%%%%

We have made the first
observation of charmless baryonic decay $B^\pm \to p \overline{p}
K^\pm$, the three-body $B^0 \to K^0 \pi^+ \pi^-$ and $B^0 \to K^0
K^+ K^-$, and see strong evidence of $B^\pm \to \eta K^\pm$ and $B^\pm
\to \eta \pi^\pm$.  We also observed the decay $B^\pm \to \omega K^\pm$
and measured the branching fractions for the decays $B \to K \pi$ and
$\pi \pi$.  We find no evidence for direct $CP$ violation in the 
observed decays.
%: $B^\pm \to K^\pm \pi^\mp$, $K^\pm \pi^0$, $K^0
%\pi^\pm$, $\pi^\pm \pi^0$, and $\omega K^\pm$.
By summer 2002 it is
anticipated that Belle will have collected 90 fb$^{-1}$ of data
providing a rich sample to continue the search for rare $B$ decays and
measure $CP$ violating effects in a variety of $B$ decay modes.

%%%%%%%%%%%%%%%%%%%%%%%%%%%%%%%%%%%%%%%%%%%%%%%%%%%%%%%%%%%%%%%%%%%%%%
\section*{References}
%%%%%%%%%%%%%%%%%%%%%%%%%%%%%%%%%%%%%%%%%%%%%%%%%%%%%%%%%%%%%%%%%%%%%%

\small

%% Introduction %%%%%%%%%%%%%%%%%%%%%%%%%%%%%%


\begin{thebibliography}{99}

\bibitem{REF:belle}
  	Belle Collaboration, A.~Abashian {\it et al.},
	KEK Progress Report 2000-4 (2000),
	to appear in Nucl. Inst. and Meth. A.

\bibitem{REF:kekb}
	KEKB B Factory Design Report, KEK Report 95-7 (1995),
	unpublished.

\bibitem{ckm}
	N. Cabibbo, Phys. Rev. Lett. {\bf 10}, 531 (1963);
	M. Kobayashi and T. Maskawa, Prog. Theor. Phys. {\bf 49}, 652
	(1973). 

%% B -> h h %%%%%%%%%%%%%%%%%%%%

\bibitem{kpitheory}
%	For theory discussions, see, for example, 
	A.~J.~Buras and R.~Fleischer, Eur. Phys. J. C {\bf 16}, 96
	(2000); 
	M.~Beneke, G.~Buchlla, M.~Neubert, and C.~T.~Sachrajda,
	Nucl. Phys. {\bf B591}, 313 (2000); 
%J.~Rosner, in {\it Lecture Notes TASI-2000} (World
%Scientific, Singapore, 2001); 
	Y.~Y.~Keum, H.~N.~Li, and A.~I.~Sanda,
	Phys. Rev. D {\bf 63}, 054008 (2001). 

%% B -> eta h %%%%%%%%%%%%%%%%%%%%%%%%%%%%%%

\bibitem{cleoetakst}
	CLEO Collaboration, S.~J.~Richichi {\it et al.},
	Phys. Rev. Lett. {\bf 85}, 520 (2000).

\bibitem{babaretapk}
	BABAR Collaboration, B.~Aubert {\it et al.}, 
	Phys. Rev. Lett. {\bf 87}, 221802 (2001).

\bibitem{rosner01}
	C.-W.~Chiang and J.~L.Rosner, hep-ph/0112285.

\bibitem{pdg}
  	Particle Data Group, D.~E.~Groom {\it et al.},
	Eur. Phys. J. C {\bf 15}, 1 (2000).

\bibitem{cbline}
	J.~E.~Gaiser {\it et al.}, Phys. Rev. D {\bf 34}, 711 (1986).

\bibitem{argus}
	H.~Albrecht {\it et al.}, Phys. Lett. {\bf B241}, 278 (1990).

%% B -> omega h %%%%%%%%%%%%%%%%%%%%

\bibitem{cleo-prl-85}
	CLEO Collaboration, C.P. Jessop {\it et al.},
	Phys. Rev. Lett. {\bf 85}, 2881 (2000).

%% B -> Ks h h %%%%%%%%%%%%%%%%%%%%%%%%%%%%%%

\bibitem{khhprd} Belle Collaboration, A. Garmash {\it et al.}, 
Phys. Rev. D {\bf 65}, 092005 (2002).


%% B -> p p K %%%%%%%%%%%%%%%%%%%%%%%%%%%%%%

\bibitem{cleobcbary} CLEO Collaboration, X. Fu {\it et al.}, 
Phys. Rev. Lett. {\bf 79}, 3125 (1997).

\bibitem{cleorrbary} CLEO Collaboration, T.E. Coan {\it et al.},
Phys. Rev. D {\bf 59}, 111101 (1999); 
CLEO Collaboration, T. Bergfeld {\it et al.}, 
Phys. Rev. Lett. {\bf 77}, 4503 (1996).

\bibitem{argusbary} 
	ARGUS Collaboration, H. Albrecht {\it et al.}, 
	Phys. Lett.  B {\bf 209}, 119 (1988).

\bibitem{delphibary} 
	DELPHI Collaboration, P. Abreu {\it et al.} 
	Phys. Lett. B {\bf 357}, 255 (1995); 
	DELPHI Collaboration, W. Adam {\it et al.} 
	Z. Phys. C {\bf 72}, 207 (1996).

\bibitem{bellebary}
	Belle Collaboration, K. Abe {\it et al.}, 
	Phys. Rev. D {\bf 65}, 091103(R) (2002). 

\bibitem{ppk}
	Belle Collaboration, K. Abe {\it et al.}, 
	Phys. Rev. Lett. {\bf 88}, 181803 (2002).

%\bibitem{Kpi}
%	CLEO Collaboration, R. Godang {\it et al.},
%	Phys. Rev. Lett. {\bf 80}, 3456 (1998);
%	Belle Collaboration, K. Abe {\it et al.}, 
%	Phys. Rev. Lett. {\bf 87}, 101801  (2001);
%	BaBar Collaboration, B. Aubert {\it et al.}, 
%	Phys. Rev. Lett. {\bf 87}, 151802 (2001)  

\bibitem{rosner_baryon} 
	G. Eilam, M. Gronau, and J. Rosner,
	Phys. Rev. D {\bf 39}, 819 (1989).

\bibitem{glueball}
	C.-K. Chua, W.-S. Hou, and S.-Y. Tsai,
	hep-ph/0204186.

\bibitem{hou_baryon} C.-K. Chua, W.-S. Hou, and S.-Y. Tsai,
	hep-ph/0204185; 
	Phys. Lett. {\bf B528}, 233 (2002).

\bibitem{cheng_baryon} H.-Y. Cheng and K.-C. Yang, hep-ph/0112245.

\bibitem{feldman} 
	G. J. Feldman and R. D. Cousins, 
	Phys. Rev. D {\bf 57}, 3873 (1998). 

%% DCPV %%%%%%%%%%%%%%%%%%%%%%%%%%%%%%

\bibitem{bib:wolfenstein}
	L. Wolfenstein, Phys. Rev. Lett. {\bf 51}, 1945 (1983).

\end{thebibliography}
\end{document}